\documentclass[superscriptaddress,amsmath,amssymb]{revtex4}
\usepackage[utf8x]{inputenc}
\usepackage[T1]{fontenc}
\usepackage{gensymb}
\usepackage{upgreek}
\usepackage{physics}              				
\usepackage{bm}
\usepackage{xcolor,graphicx}
\usepackage{amsmath,amsfonts,amssymb}
\usepackage{comment}
\usepackage{enumitem}
\usepackage{array}
\usepackage{graphicx}
\usepackage[colorinlistoftodos]{todonotes}
\usepackage[colorlinks=true, citecolor=red, linkcolor=blue, urlcolor=blue ]{hyperref}
\usepackage{cleveref}
\usepackage{float}
\usepackage{wrapfig}
\usepackage{setspace}
\usepackage{natbib}
\usepackage{graphicx}
\usepackage{dcolumn}
\usepackage{bm} 
\usepackage{siunitx}
\usepackage{filecontents}
\usepackage{mathrsfs}

\newcommand*{\doi}[1]{\href{http://dx.doi.org/#1}{doi: #1}}

\begin{document}

\title{Rebound and scattering of motile {\it Chlamydomonas} algae in confined chambers}

\author{Albane Th\'ery}
\email{at830@cam.ac.uk}
\affiliation{Department of Applied Mathematics and Theoretical Physics, University of Cambridge, Cambridge CB3 0WA, United Kingdom.}
  
\author{Yuxuan Wang}
 \affiliation{Universit\'e de Paris, CNRS UMR 7057, Laboratoire Mati\`ere et Syst\`emes Complexes MSC, F-75006 Paris, France.}
\author{Mariia Dvoriashyna}
 \affiliation{Department of Applied Mathematics and Theoretical Physics, University of Cambridge, Cambridge CB3 0WA, United Kingdom.}
 \author{Christophe Eloy}
 \affiliation{Aix Marseille Univ, CNRS, Centrale Marseille, IRPHE, 13013 Marseille, France. }
 \author{Florence Elias}
 \affiliation{Universit\'e de Paris, CNRS UMR 7057, Laboratoire Mati\`ere et Syst\`emes Complexes MSC, F-75006 Paris, France.}
\author{Eric Lauga}
 \email{e.lauga@damtp.cam.ac.uk}
 \affiliation{Department of Applied Mathematics and Theoretical Physics, University of Cambridge, Cambridge CB3 0WA, United Kingdom.}

\begin{abstract}
    Motivated by recent experiments demonstrating that  motile algae get trapped in draining foams, we study the trajectories of microorganisms confined in model foam channels (section of a Plateau border). We track single {\it Chlamydomonas reinhardtii}  cells confined in a thin three-circle microfluidic chamber and show that  their spatial distribution exhibits strong corner accumulation. Using  empirical scattering laws observed in   previous experiments (scattering with a constant scattering angle), we next develop a two-dimension geometrical model and compute the phase space of trapped and periodic trajectories of swimmers  inside a three-circles billiard. We find that the majority of cell trajectories end up in a corner, providing a geometrical mechanism for corner accumulation. Incorporating the distribution of scattering angles observed in our experiments and including hydrodynamic interactions between the cells and the surfaces into the geometrical model enables us to reproduce the experimental probability density function of micro-swimmers in microfluidic chambers. Both our experiments and models demonstrate therefore that motility leads generically  to trapping in complex geometries.
\end{abstract}

\maketitle


\section{Introduction}

Microorganisms display a wide range of mechanisms in order to swim in fluids on small scales~\cite{brennen77,lauga2009hydrodynamics,elgeti2015physics}. 
Although the flagellar structure is a highly conserved trait in both prokaryotic and eukaryotic motile cells, the detailed mechanisms for self-propulsion differ~\cite{braybook}.  Most bacteria use the passive rotation of helical flagellar filaments by a basal rotary motor~\cite{berg1973bacteria,silverman1974flagellar}. In contrast, eukaryotic cells swim using  active flexible flagella that bend in periodic waves through the coordinated contraction of distributed  motor proteins~\cite{lindemann1994model, nicastro2006molecular}.  Mammalian spermatozoa are propelled by a single flagellum~\cite{ishijima1986flagellar}, while the model green algal genus {\it Chlamydomonas} has two flagella \cite{harris2001chlamydomonas} and ciliates, such as the model genus {\it Paramecium}, are actuated by an array of small synchronised flagella termed  cilia~\cite{blake74}.

These diverse swimming mechanisms have been extensively studied, in particular those of model microorganisms    with well-established experimental protocols, such as the bacterium {\it Escherichia coli} ({\it E.~coli})~\cite{berg2008coli} or the alga {\it Chlamydomonas reinhardtii} (CR)~\cite{harris1989chlamydomonas,goldstein2015green}. Along with experimental work, significant  theoretical modelling has been developed  to quantify and predict the motion of swimming cells~\cite{brennen77,lauga2009hydrodynamics,elgeti2015physics,lauga_book}.  In particular, the flow generated by the breaststroke motion of  free-swimming CR cells used in this paper has been quantified experimentally and modelled in the far field using three point forces for the body and two flagella~\cite{ruffer1985high, polin2009chlamydomonas,drescher2010direct,guasto2010oscillatory, tam2011optimal}.
  
The habitat of microorganisms is often far from the idealised bulk fluids considered in model biophysical experiments and theoretical modelling. Cells interact with their surroundings through complex hydrodynamic and steric interactions~\cite{drescher2011fluid,jakuszeit2019diffusion}. Rigid boundaries, free surfaces and other obstacles in suspensions are known to strongly affect  swimming behaviour~\cite{bechinger2016active,takagi2014hydrodynamic}. Relevant complex media include thin films~\cite{guasto2010oscillatory}, porous media such as soil~\cite{theves2015random}, the ovary tract in the case of mammalian spermatozoa~\cite{denissenko2012human}, or tissues within higher  organisms, including the bloodstream~\cite{heddergott2012trypanosome} and gastrointestinal mucus for pathogens~\cite{ottemann1997roles}.   Complex environments are therefore ubiquitous  in the life of   self-propelled microswimmers, and as a result their swimming patterns have been studied both theoretically and experimentally 
 in a wide range of geometries, including near rigid surfaces~\cite{rothschild_1963,berke2008hydrodynamic}, corners~\cite{shum2015hydrodynamic,nosrati2016predominance}, in strong confinement~\cite{mondal2020inversion}, in the interstices between inclined plates~\cite{ishikawa2019swimming}, in channels~\cite{zottl2013periodic}, as well as in droplets and crowded environments~\cite{spagnolie2015geometric,contino2015microalgae,chepizhko2019ideal}. 

\begin{figure*}
\begin{center}
\includegraphics[width=16.5cm]{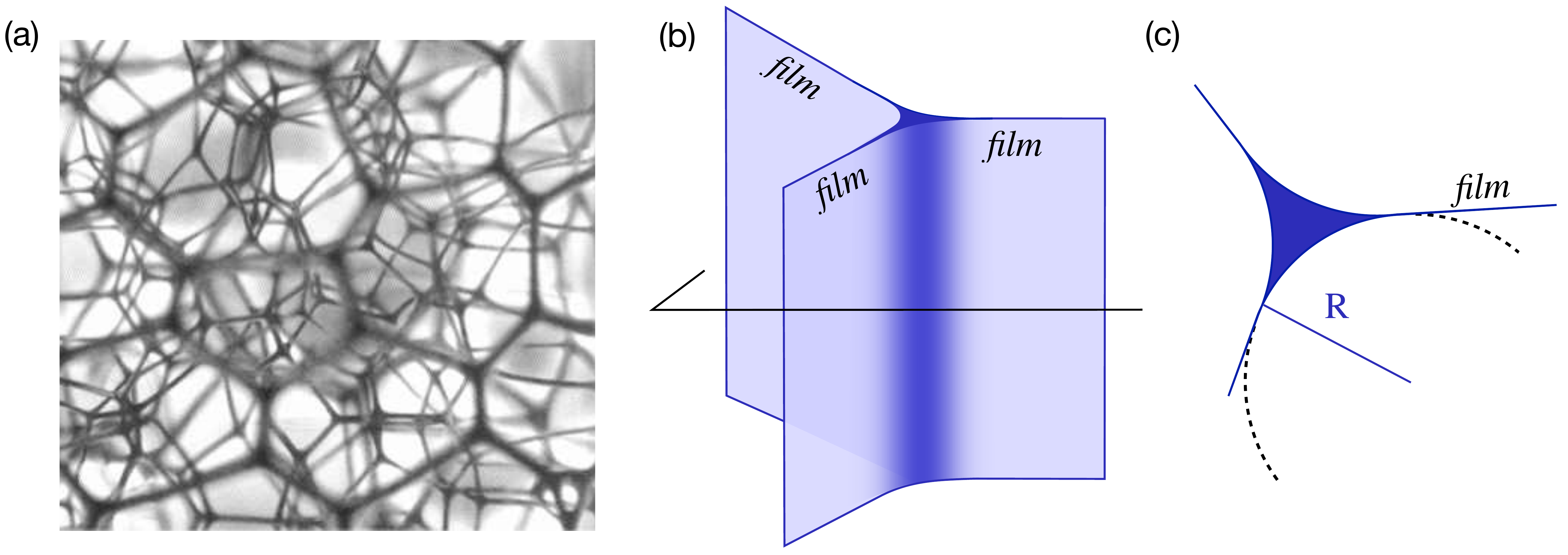}
\caption{Local structure of a foam. (a) Image of the internal foam liquid skeleton with the liquid channels, called Plateau borders, appearing in black (Image credit: own work);  (b) Structure of a single Plateau border, at the intersection between three soap films making equal angles of $2 \pi / 3$; (c) Cross-section of a Plateau border, with three concave edges, each of constant curvature $R$. 
}
\label{Fig_foam}
\end{center}
\end{figure*}

Long-range hydrodynamic interactions between  microswimmers and    any obstacles surrounding them are governed by the flow created by  the beating flagella and the motion of the cell body~\cite{boddeker2020dynamic}. Swimming organisms are typically divided into two categories, depending on the type of long-ranged flow they create in the surrounding fluid: pushers, a group that includes spermatozoa and {\it  E.~coli} bacteria~\cite{berke2008hydrodynamic}, are pushed from the  back by their flagella; in contrast, pullers, such as  CR cells~\cite{drescher2010direct}, swim flagella first. This distinction originates from the  symmetries in the flow field created by the swimmers, with a puller being equivalent to a pusher seen under a time-reversal symmetry, and as a  result these two categories lead to very different dynamic behaviours close to boundaries and obstacles. Noise~\cite{schaar2015detention} and steric interactions~\cite{kantsler2013ciliary,contino2015microalgae} also affect the trapping and scattering of swimmers of surfaces, including CR in microfluidic chambers~\cite{ostapenko2018curvature}.

An important instance of swimming in a complex environment is the propagation of microorganisms in porous media such as soils  \cite{ranjard2001quantitative} and foams~\cite{roveillo2020}. The formation of aquatic foams on certain rivers, lakes and coastlines has been reported to be associated with a loss of phytoplanktonic biomass in the water column~\cite{seuront2006, schilling2011}. The role of aquatic foams on microorganism populations has, however, yet to be understood. The trapping of algae brings them out of their aquatic environment, thus leading to depletion, but this could also favour dissemination to new environments. Exploring the physical mechanisms at play in the interactions of motile organisms with foams would strengthen our understanding of the influence of foam formation on local ecosystems. 

Recent work in this direction investigated the fate of planktonic biomass trapped in foams, showing experimentally that flagellated CR cells  remain actively trapped over long periods of time in a draining foam, while passive bodies of the same size and density (including dead cells) escape the foam  with the  gravity-driven flow~\cite{roveillo2020}.   The liquid part of a foam consists of interconnected micro-channels formed by the edges of contacting bubbles, in which the liquid flows as in a pipe. These microchannels, called Plateau borders, have a well-defined structure imposed by interfacial minimisation and  are classically described by Plateau's rules~\cite{cantat2013} wherein bubbles always meet by three. The cross-section of the Plateau border is therefore triangular with concave curved sides (see illustration in Fig.~\ref{Fig_foam}).  
Microscopic observations of CR cells swimming  in a chamber mimicking the cross-section of foam Plateau border further revealed that cells accumulate near the corners of the Plateau border~\cite{roveillo2020}. We note that the presence of surfactants rigidifies  the  foam boundaries in the experiments, thus allowing comparison with no-slip walls~\cite{roveillo2020}.

In this article, we present a  combined experimental and  theoretical analysis of the experimental swimming behaviour of CR cells in two-dimensional (2D) microscopic chambers imitating the cross-section of a single foam Plateau border. We first use  tracking  data on the  swimming  dynamics of the  cells in three different chambers to derive the full steady-state probability distribution function (pdf)  of swimming CR  cells. This distribution is seen to be strongly peaked in the corners  of the chambers. Next, we use 
  empirical scattering laws observed in   previous experiments (scattering with a constant scattering angle) to   analyse theoretically the phase space of trapped and escaping trajectories inside a three-circles billiard. We show that  the experimentally-observed accumulation of swimmers in the corners has a geometrical origin. We then develop a more detailed theoretical model based on experimental data to  reproduce quantitatively the probability density function of swimmers in the chambers. We show that the trapping of CR is controlled by the shape of the concave microchambers, the finite size of the CR cells, and the angle distribution of the cells scattering off the walls. We also observe that the microswimmers are significantly slowed down in the vicinity of walls at distances much larger than those of contact interactions, an effect  which, as we quantify using 
numerical simulations, is of purely hydrodynamic origin.

 \begin{figure}[b]
    \centering
    \includegraphics[width =0.8\columnwidth]{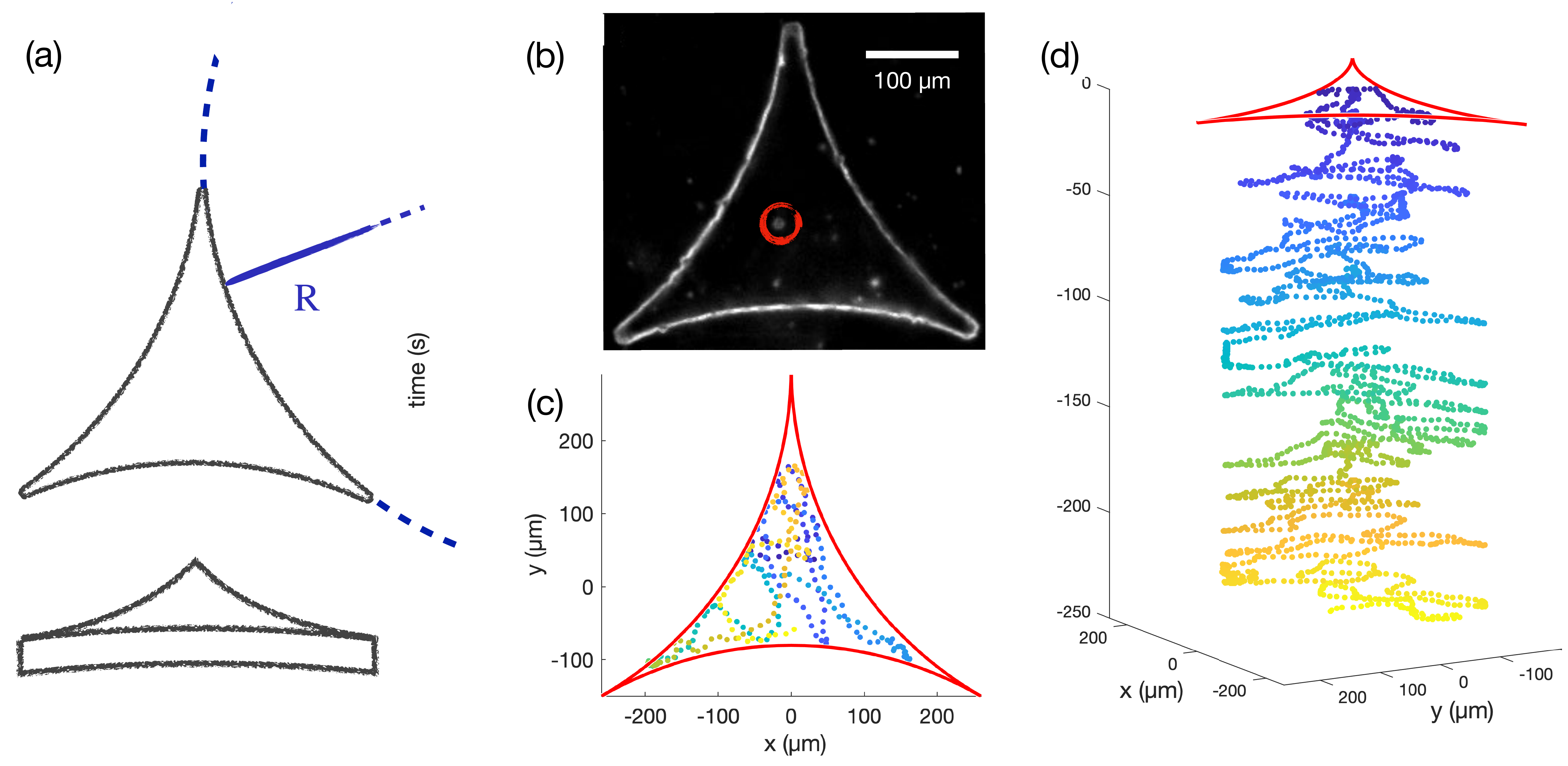}
    \caption{ (a)  Sketch of the chamber shape (top view and side view).  (b) Experimental setup: PDMS micro-chamber mimicking the cross-section of a foam channel with a single CR organism (circled). (c) Swimming trajectory of the organism for $30~\si{s}$; (d) Tracking of the  CR cell for  $250~\si{s}$ with time shown along the $z$ axis.}
    \label{setup}
\end{figure}

\section{Probability density function of cells in a quasi-two-dimensional chamber}

\subsection{Experimental set-up} 

Individual microswimmers cells were incorporated in microscopic chambers mimicking the cross-section of a foam channel, as shown on Fig.~\ref{setup}a. The chambers consists in a triangular shape with concave curved sides with identical radius of curvature $R$. The thickness of the chambers is $20~\si{\mu m}$. 
The experimental procedure described below has been presented in a previous publication~\cite{roveillo2020}. Here, we use the raw experimental data and develop a new tracking algorithm which follows the position of an individual cell and its interactions with the chamber walls.

The microswimmer used is the model alga \textit{Chlamydomonas reinhardtii} (CR), which is a single-cell green alga about $10 ~\si{\mu m}$ in diameter that swims with two flagella at the front of the organism. 
We used the wild type strain CC124- obtained from the Chlamydomonas Resource Center. The algae were inoculated into High Salt Acetate (HSA) culture medium \cite{sueoka1960} and maintained at 25 degrees under constant gentle agitation and a day/night illumination conditions of 12h/12h. The experiments were carried out between 48 and 72 hours after inoculation. This preparation ensures the synchronisation of the algae and thus the reproducibility of their behaviour.

The micro-chambers were designed using soft lithography techniques. Polydimethylsiloxane (PDMS) chips were obtained from a mould consisting in an array of identical micro-chambers. Three different sizes of PDMS microchambers were used, with $R$ = 260~$\mu$m,  520~$\mu$m and  1040~$\mu$m.
Just before any experiment, the PDMS device was made hydrophilic by oxidation using air plasma device over 2 minutes. 

Immediately after oxidation, a $\sim$ 3~$\mu$L drop of suspension of CR algae in their culture medium was deposited on the PDMS device and a microscope slide was placed gently on the top. The chambers were placed under an inverted phase contrast microscope (Olympus IX73, fitted with the contrast module PH1U and two objectives: UPLFLN4X and UCPLFLN20X). The  position of each CR cell was recorded during 5~min at a rate of 10~fps, using a red illumination (wavelength $> 630\si{.nm}$) to avoid any phototactic effect (see Fig.~\ref{setup}b). Each experiment was repeated 15 to 17 times. 
 Since we are interested in trapping in foams, which occurs at very low density and in order to avoid algae  interactions interfering with the single particle tracking, 
chambers containing one cell were used for the experiments. We thus excluded the data from 14 experiments for which two cells or more were present in the chamber.  Having multiple cells in a given experiment happened more frequently in the larger chambers, as the filling solution had constant algae concentration.
In total, we analyse 33 trajectories, 17 for the $R = 260$~$\mu$m chamber, 9 for $R =520$~$\mu$m and 7 for $R =1040$~$\mu$m. We show videos of experiments in each of the three chambers in the supplementary movies \texttt{SM2.mp4}, \texttt{SM3.mp4} and \texttt{SM4.mp4}.

We track the position of each CR cell in the chamber by subtracting the first image of the sequence and finding the local intensity maximum, which corresponds to the algae body. We check for errors in the points by defining a maximum speed of $260~\si{\mu m .s^{-1} }$. If a single point is out of this speed scope, we replace it by the average of the previous and subsequent points. If more than one point is wrong, we move to manual tracking for this set of images. Obtained trajectories lasting  $30~\si{s}$ and $250~\si{s}$, respectively,  are depicted in Figs.~\ref{setup}c and d, with the colormap indicating the direction of time, from dark blue to yellow. For clarity, the $z$-direction of Fig.~\ref{setup}d also represents time. Supplementary video \texttt{SM1.mp4} also shows tracking in a $R =520$~$\mu$m chamber.

\subsection{Probability density function}

 \begin{figure*}[t]
    \centering
    \includegraphics[width = 0.95\textwidth]{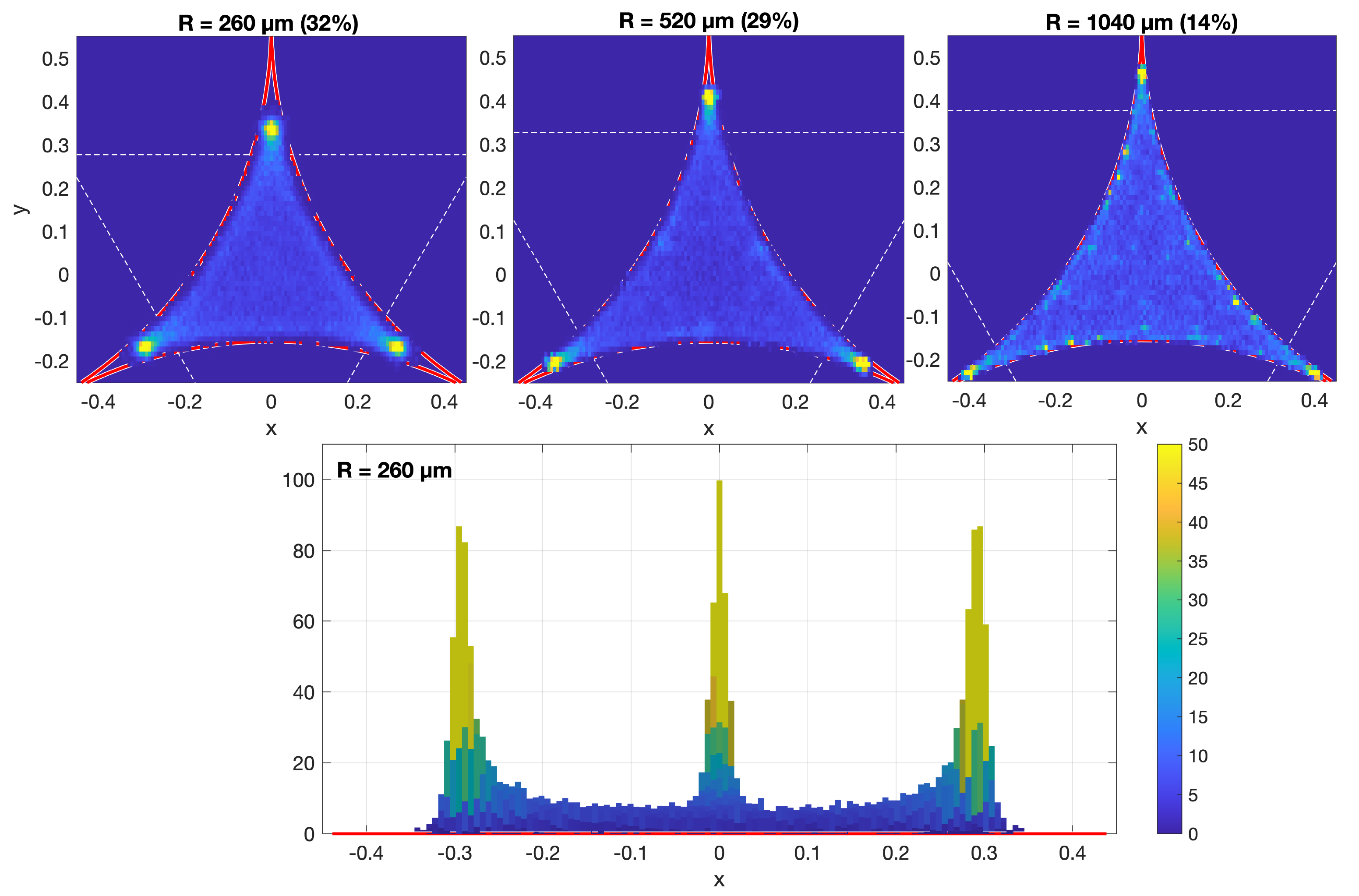}
    \caption{Probability density function for the location of the swimming cells in the three chambers with different wall radii $R$ ($260$~$\mu$m, $520$~$\mu$m and $1040$~$\mu$m), nondimensionalised using $R$ for unit length and using bins of size $1/150$. Text in parenthesis:  percentage of time spent in corners is given for each chamber. The red circles represent the  theoretical circles of radius 1, and the white dashed lines show the corner domains used to quantify corner accumulation  (which depends on corner sharpness of the chamber). The bottom panel is a side view of the distribution in the small chamber.}
    \label{pdfexp}
\end{figure*}

The   positions of each cell   were statistically averaged over all experiments in a given chamber and converted into relative probability density maps for the algae position. Additional averages were performed using rotation of the entire map by angles $ 2 \pi /{3}$ and  $4 \pi / 3$, thereby exploiting   the angular symmetry of the Plateau border cross-section. These pdfs are presented in Fig.~\ref{pdfexp} for the three different chambers, with both colour and height (bottom panel: side view) representing the magnitude of the probability density function for the swimmer position. Lengths are nondimensionalised using the radius $R$ of the three walls as unit. 

The most striking feature of the pdfs is a sharp accumulation of the swimmers in the corners of the Plateau borders. The models developed in the following sections of the paper will   explain this phenomenon of corner accumulation, a key element to understanding the trapping of motile cells in complex geometries such as foams.

We measure that cells spend about a third of their time in corners, with details varying with the chamber (and therefore the sharpness of the corners); the  exact fraction of time spent in the corners is shown in Fig.~\ref{pdfexp} (top). Inspecting individual trajectories where algae swim into a corner, we see that after some time the trapped cells turn back and escape. These cells then follow a new trajectory and eventually end up stuck in another corner, and the process repeats. Those half-turn motions of the trapped cells will be explicitly taken into account in Section~\ref{sec:quant}.

A second feature of the algae trajectories is that the cells spend more time swimming near boundaries than in the centre of the chamber. Contact interaction of CR with walls has been previously studied~\cite{kantsler2013ciliary}, and the authors showed that cells have a tendency to swim along the boundary before scattering away. Wall accumulation of CR has also been reported in spherical and elliptical channels, which has been rationalised using an active Brownian 
model~\cite{ostapenko2018curvature}. By measuring the range of the interaction, we show below that, while contact interactions  dominate surface interactions~ \cite{kantsler2013ciliary}, far-field hydrodynamic effects must also be taken into account in order to reproduce quantitatively the distribution of swimming cells in chambers. We will also show that  corner accumulation is not due to boundary-following swimming but that instead it has a geometrical origin. 

 \begin{figure}[b]
    \centering
    \includegraphics[width = 0.7\columnwidth]{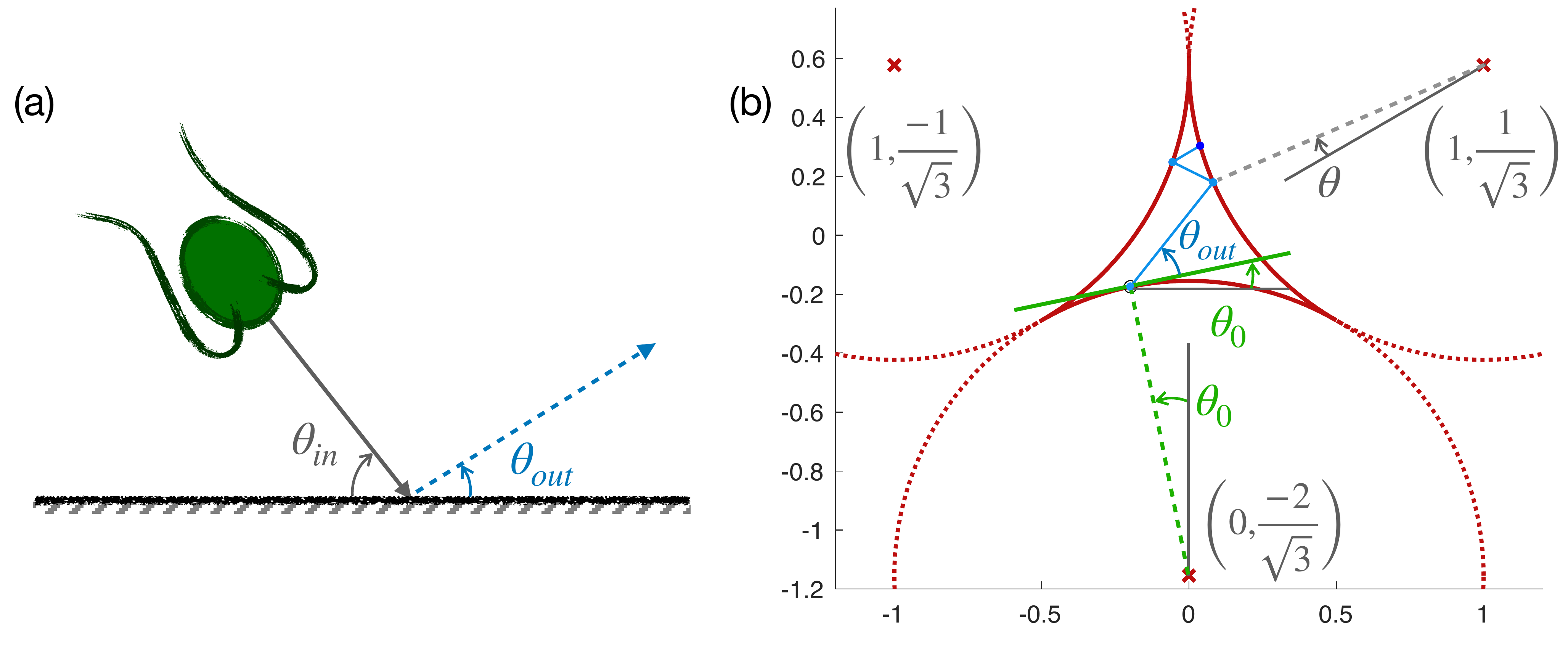}
    \caption{Parameters and geometry of the system. (a) Individual scattering event of a swimmer, with incident angle $\theta_{in}$ and scattering angle $\theta_{out}$. (b) Geometry of the chamber with the three circles of radius $R=1$, the position of their centres and an example trajectory with the constant scattering angle $\theta_{out} = 0.7~\si{\radian} $. The trajectory starts on the bottom circle at the initial angle $\theta_0 = 0.2~\si{.\radian} $, coming from the left. The consecutive contact points are characterised by the circle on which they occur, their position measured by $\theta$ and the side of the normal from which the swimmer arrives.}
    \label{systgeom}
\end{figure}

In Fig.~\ref{pdfexp}, we further observe some smaller peaks in the pdf close to some walls, especially in the $1040 ~\si{\mu m}$ chamber (light colours). These correspond to experimental irregularities or dust particles close to the chamber walls, which then decrease the time spent in corners by the cells; they are not relevant for our analysis in the context of foams and wall interactions and can thus be ignored in our modelling approach.

\section{A geometrical approach to  trapping in corners}

\label{geomodel}
\subsection{Two-dimensional model for algae scattering with constant scattering angle}

In this section, we develop a simple  geometrical model to explain the most striking aspect of the algae spatial distribution, i.e.~the  accumulation of the swimming cells in the chamber corners.
 We take a two-dimensional geometrical approach to the problem and consider a cell swimming  in a domain obtained as the outside of the three tangent circles (Fig.~\ref{systgeom}). We study mathematically all  possible trajectories of the cell and show that most of them end in a corner.

 We assume that the cells swim along straight lines before scattering off   walls. Experimentally, the swimming behaviour of CR cells is ballistic at short times and diffusive at long times~\cite{kantsler2013ciliary} with a transition time of several seconds. When the cells are confined in a plane, however, the trajectories become predominantly ballistic~\cite{ostapenko2018curvature}, justifying this modelling assumption.

A key element is the modelling of individual scattering events when the swimmer encounters a wall. We assume each scattering event to be punctual, and  neglect here any sliding of the algae along the chamber wall (which will be incorporated in the detail model of Section~\ref{sec:quant}). Motivated by past  experimental data~\cite{kantsler2013ciliary}, we assume here that the orientation of the swimmer after the wall-bounce  is  independent of its initial orientation. Specifically, if we denote, for a given scattering event, the incident angle of a CR cell $\theta_{in}$  and the scattering angle $\theta_{out}$  (see notation Fig.~\ref{systgeom}a), we assume 
$\theta_{out}$ to be independent of $\theta_{in}$.

Additionally, we take in this section the scattering angle $\theta_{out}$ to be fixed and constant for a given swimmer. This choice for $\theta_{out}$ is an important difference with well-studied classical billiards~\cite{chernov2006chaotic}, including three-disks ones~\cite{sano1994bifurcation,weibert2002periodic}, where the particle reflection is specular, meaning that $\theta_{in} = \theta_{out}$. Microorganisms billiards with constant scattering angle $\theta_{out}$, also motivated by experimental  work~\cite{kantsler2013ciliary}, have been previously studied in polygonal geometries~\cite{spagnolie2015geometric} as well as ellipses and more complex closed curves~\cite{krieger2016microorganism}. We extend here these analysis to the particular geometry of the Plateau border section, thus providing a link between past work on   microorganisms billiards and our new experiments.

Because of the existence of cusps  in our billiard geometry (corner points with zero interior angles), `trapping' events in which the swimmer never exits a corner can occur. Note that this would also be true in geometries with acute angles, such as triangles. In this paper we denote `escaping' the  opposite of trapping, keeping in mind our initial motivation of understanding the fate of algae in foams. We are interested in what follows in  characterising the possibility of trapping events, and in quantifying how likely they are to occur as a function of  the value  of $\theta_{out}$. We demonstrate that the concavity of the walls makes the future of a given trajectory (trapped vs escape)  dependent on the initial position of the swimmer, which is not the case for polygons. Consequently, in this section, we aim to determine the phase space of trapped trajectories as a function of the  initial conditions of the cell and on the (constant) scattering angle.

\subsection{System geometry}

In order to list all the possible trajectories, we parametrise our system as shown in Fig.~\ref{systgeom}. 
A trajectory is characterised by  the value of $\theta_{out}$ and the initial position and  swimming direction of the cell. Instead of listing all possible starting points and orientations in the chamber, we may only consider the first contact point with the walls and the side of the normal from which the swimmer arrived. 
Taking into account the angular symmetry of the three circles for rotations of $2 \pi /3$ and $4 \pi / 3$, we can limit our analysis to trajectories starting from the bottom circle.  
The first contact point is defined by the angle $\theta_0$ measured from the centre of the bottom circle and taking its value between $-\pi / {6}$ (left corner) and  $\pi / {6}$ (right corner) (see illustration in Fig.~\ref{systgeom}a). In what follows, all angles defining the position of the swimmer on the circles are oriented counter-clockwise. Further, for a given contact point, the swimmer comes from either left or right side of the normal. Recalling the mirror symmetry between $\theta_0$ and $-\theta_0$, we only consider trajectories coming from the left at the first contact point. For consistency with past experiments~\cite{kantsler2013ciliary} and previous analysis of microswimmers billiards~\cite{spagnolie2017microorganism}, the model swimmer is assumed to always leave from the opposite side of the normal, which means that $\theta_{out} \in [0; \pi / {2}]$. Possible reversals of the swimming direction upon scattering off a wall will be included in the following section where non-constant values of $\theta_{out}$ are discussed.

With the appropriate symmetries, each possible swimmer trajectory in the geometrical approach is therefore indexed by a single ($\theta_0$,$\theta_{out}$) pair, with  $\theta_{0} \in [-\pi /{6} ; \pi / {6}]$ and  $\theta_{out} \in [0; \pi / {2}]$, and the trajectory of the swimmer is  fully deterministic. 
The consecutive contact points with the walls are characterised by a position measured by the angle $\theta \in [-\pi / {6}; \pi / {6}]$ on one of the three circles  (see Fig.~\ref{systgeom}a) , and by the direction of arrival of the swimmer with respect to the normal at the contact point. We investigate the outcome of all the possible trajectories, either trapped in a corner or escaped, for each pair of angles ($\theta_0$,$\theta_{out}$).

 \begin{figure*}[t]
    \centering
    \includegraphics[width = 0.9\textwidth]{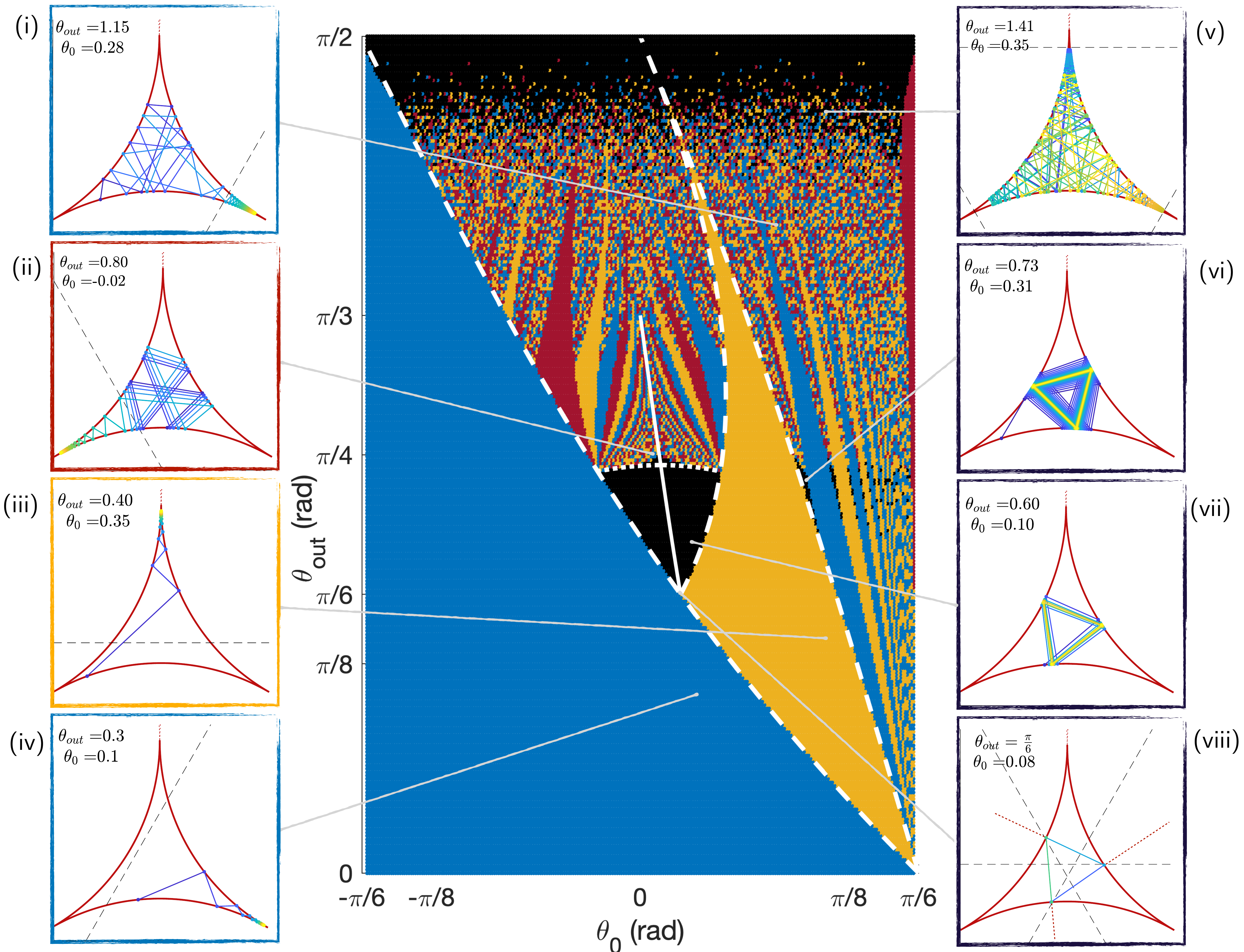}
    \caption{Phase map of the trapping vs escaping outcome for swimmers undergoing  200 bounces with different initial positions $\theta_0$ and scattering angles $\theta_{out}$. These results are obtained numerically. 
    The escaping trajectories are shown in black and are illustrated on the right panels (v to viii). The trapping trajectories  are shown in colours, with illustrations on the left panels: blue for the right corner of Fig.~\ref{systgeom}i and iv, yellow for the top corner (iii), and red for the left one (ii). The white lines correspond to analytical limits for some of the regions of the phase map: the plain line depicts periodic triangular orbits, the dotted line six-points periods, and the dashed lines some coloured regions areas. In the panels, the colour of the trajectory  indicates the swimming direction (increasing time from blue to yellow), and the dashed black line shows the attraction range of the corner region, as detailed in Section~\ref{trapped}.}
    \label{mapexit}
\end{figure*}

\subsection{Phase map of the trajectories}

By integrating the model  numerically, we can obtain the outcome of all possible trajectories indexed by the ($\theta_0$, $\theta_{out}$) pairs and look at whether they end up in a corner. To do so, we run simulations for 250 evenly spaced values for both $\theta_0 \in [-\pi/6 + 0.05 ; \pi/6 - 0.05]$ and $\theta_{out} \in [0; \pi/2]$. We stop each simulation after 200 bounces and report on  the position of the swimmer in the chamber. We plot the results for all trajectories in the phase map of Fig~\ref{mapexit}. There are four different possible outcomes for each ($\theta_0$, $\theta_{out}$), as illustrated using four colours: no trapping after 200 reflections (black) trapping in right (blue), top (yellow) and left (red) corners. Illustrative trajectories are also shown on the figure, with trapped trajectories on the left and escaping ones on the right.

The most striking feature of this phase map is that a large majority of the trajectories, over $80 \%$, end in one of the three corners. For small values of $\theta_{out}$ or $\theta_0$, all trajectories end up in the right corner from the first bounce (blue colour in Fig.~\ref{mapexit}), a result true for nearly half of all swimmers. 
This is due to the existence of regions of the chamber around the corners from which the swimmer cannot escape. 
We show later that the size of these trapping regions decreases with $\theta_{out}$, leading to the appearance of more intricate trajectories before trapping and even escaping ones at high scattering angles. Indeed, for large values of $\theta_{out}$, the structure of the phase space becomes more complex. We further note that there are two regions where some trajectories escape the corners (these are the  black areas on the phase map): some  orbits with intermediate values of $\theta_{out}$ are perfectly periodic, and some trajectories come close to the  corners but escape them as $\theta_{out} \rightarrow \pi / {2}$. 
In the rest of this section, we interpret the main areas of the phase space and obtain some analytical expressions for their boundaries; these analytical limits are depicted as white lines in Fig.~\ref{mapexit}.

\subsection{Understanding the phase map}

\subsubsection{Trapped trajectories}

\label{trapped}

\begin{figure}[t]
    \centering
    \includegraphics[width = 0.85\columnwidth]{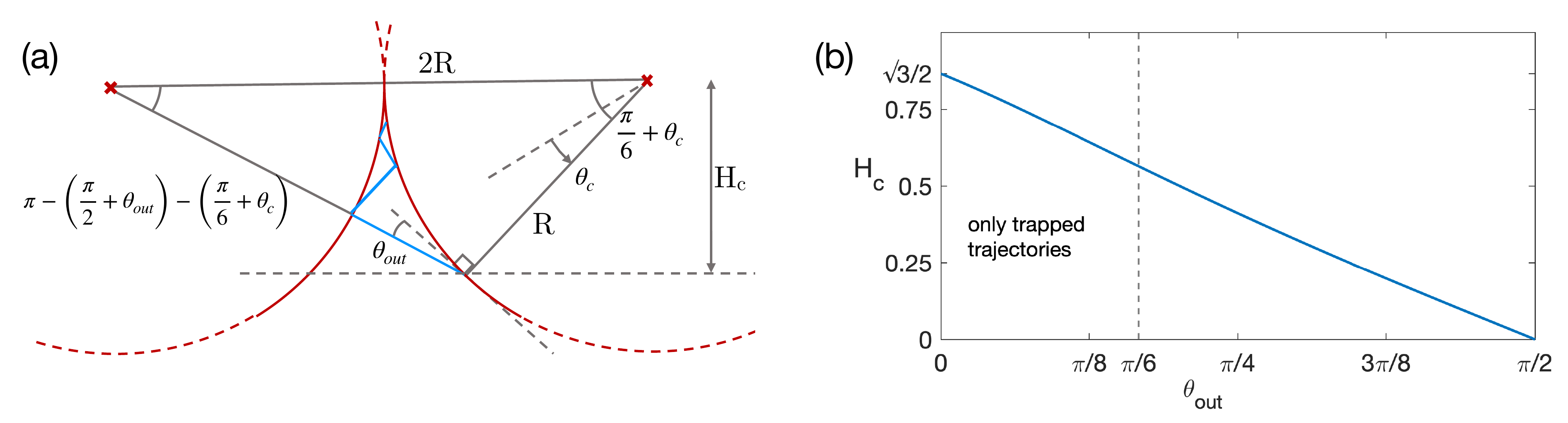}
    \caption{(a) Above the dashed line is the trapping region: if the swimmers enters this region, it will swim into the corner.  (b) Size of the trapping region $H_c$ (defined in (a)) as a function of the scattering angle $\theta_{out}$.  The vertical dashed line at $\theta_{out} = \pi/6 $ shows the limit scattering angle below which all trajectories are trapped }
    \label{critr}
\end{figure}

Close to each corner, there is a critical trapping region where all entering swimmers end up in the corner. This region is depicted with a dashed black lines on trajectories (i) to (iv) of Fig.~\ref{mapexit}. The limit trajectory entering this region is the one where the swimmer reaches a wall along the normal to the contact point, as shown in Fig.~\ref{critr}. Using the law of sines in the triangle shown in Fig.~\ref{critr}a, we can define the critical contact angle $\theta_c$ in the following way: 

\begin{equation}
 \frac{\sin \left[ \frac{\pi}{2}+\theta_{out} \right]}{2R} = \frac{ \sin \left[ \pi - ( \frac{\pi}{2}+\theta_{out}) - (\frac{\pi}{6} + \theta _c) \right] }{ R}
. \end{equation}
A  swimmer coming from outside this region is therefore sure to be trapped in the corner when it touches a wall with $ |\theta| > \theta_{c}$, with 
 \begin{equation}
    \theta_{c} = \arccos \left[\frac{\cos(\theta_{out})}{2} \right]- \frac{\pi}{6} - \theta_{out}  .
\label{eqtc}
\end{equation}
 We note that a trajectory might leave this region if it starts inside it but is oriented outwards. 
We also compute the extension of the trapping region $H_c$ defined as the distance between the corner and the boundary of the trapping region, as shown on Fig.~\ref{critr}a. $H_c$ is the height of the previously used triangle, with
\begin{equation}
     \sin\left(\frac{\pi}{6} + \theta_{c}\right) = \frac{H_{c}}{R},
\end{equation}
which we rewrite using Eq.~(\ref{eqtc}) as
\begin{equation}
    H_{c} = R \sin \left[ \arccos \left( \frac{\cos(\theta_{out})}{2} \right) - \theta_{out} \right]
\label{eqh}.
\end{equation}
  The dependence of the size of the trapping region on $\theta_{out}$ is plotted Fig.~\ref{critr}b 
 and we see a systematic decrease, with $\theta_{out}$, from $H_c=R\sqrt{3} /{2}$ at zero scattering angle to 0 when $\theta_{out} = \pi / {2}$.

When the sizes of the three trapping regions are made to increase (by lowering the value of $\theta_{out}$), we can reach a point where all three  cover the whole chamber; this occurs for $H_c = R/{\sqrt{3}}$  and $\theta_{out} \approx 0.35$. 
We therefore know that there exists a limit value for $\theta_{out}$, equal to or larger than $0.35$,  below which all swimmers end up trapped, as  seen on the phase map in Fig.~\ref{mapexit}. The limiting case corresponds to the lowest value of $\theta_{out}$ for which the swimmer can successfully escape all three corners. In this trajectory, the swimmer reaches all three circles precisely at the border of the trapping regions, as presented on Fig.~\ref{mapexit}viii. This occurs for $\theta_{out} = \pi / {6}$, which is the desired limit.
In that case, all trajectories arrive normally and are therefore aligned with the three centres of the circles. 

We note that  experimental work on the scattering angles~\cite{kantsler2013ciliary} found that the distribution of $\theta_{out}$ exhibits a peak at $\theta_{out} \approx 0.28~\si{ \radian}$. Using this value in our   model  results in prediction of trapping for all swimmers with any  initial position. 

The existence of a trapping region due to the cusps in the geometry considered in our paper is an important difference with the case of a polygonal chamber, and in particular a triangular one. It means that attractive periodic orbits, if they exist, only have a limited attraction range, making trapping more likely.

\subsubsection{Periodic orbits}

We now focus on a first type of escaping trajectories, namely periodic orbits. These periodic trajectories can be attractive, so that trajectories starting at a close enough value of $\theta_0$ will tend to the periodic one, and are thus escaping trajectories as well. The only periodic trajectories that lead to a non-zero probability of escape for a given value of $\theta_{out}$ are the attractive ones. 
As we are interested in the geometrical mechanisms for corner accumulation, we now identify some of these periodic orbits, in particular  the  attraction basin of the  attractive ones.

We therefore look for periodic values $(\theta_0^*, \theta_{out}^*)$ as well as neighbouring trajectories $(\theta_0^* + \delta \theta_0, \theta_{out}^*)$ tending to the periodic trajectory (as in Fig.~\ref{mapexit}vi and vii). Since periodic orbits are, by definition, trajectories that reach the same point after a number of bounces, we look for a  mathematical relationship between the position of consecutive contact points. A swimmer starting from a circle can reach either of the two remaining walls, and its scattering orientation depends on the side of the normal from which it arrives at the wall;  there is   unfortunately no  general analytical expression for the position successive contact points of a trajectory. However, we can find expressions relating consecutive scattering angles in some specific cases. This enables us to express the position of some of the regions on the phase map, in particular some positions for periodic $(\theta_0^*, \theta_{out}^*)$ and attraction range. 

\begin{figure}[t]
    \centering
    \includegraphics[width = 0.8\columnwidth]{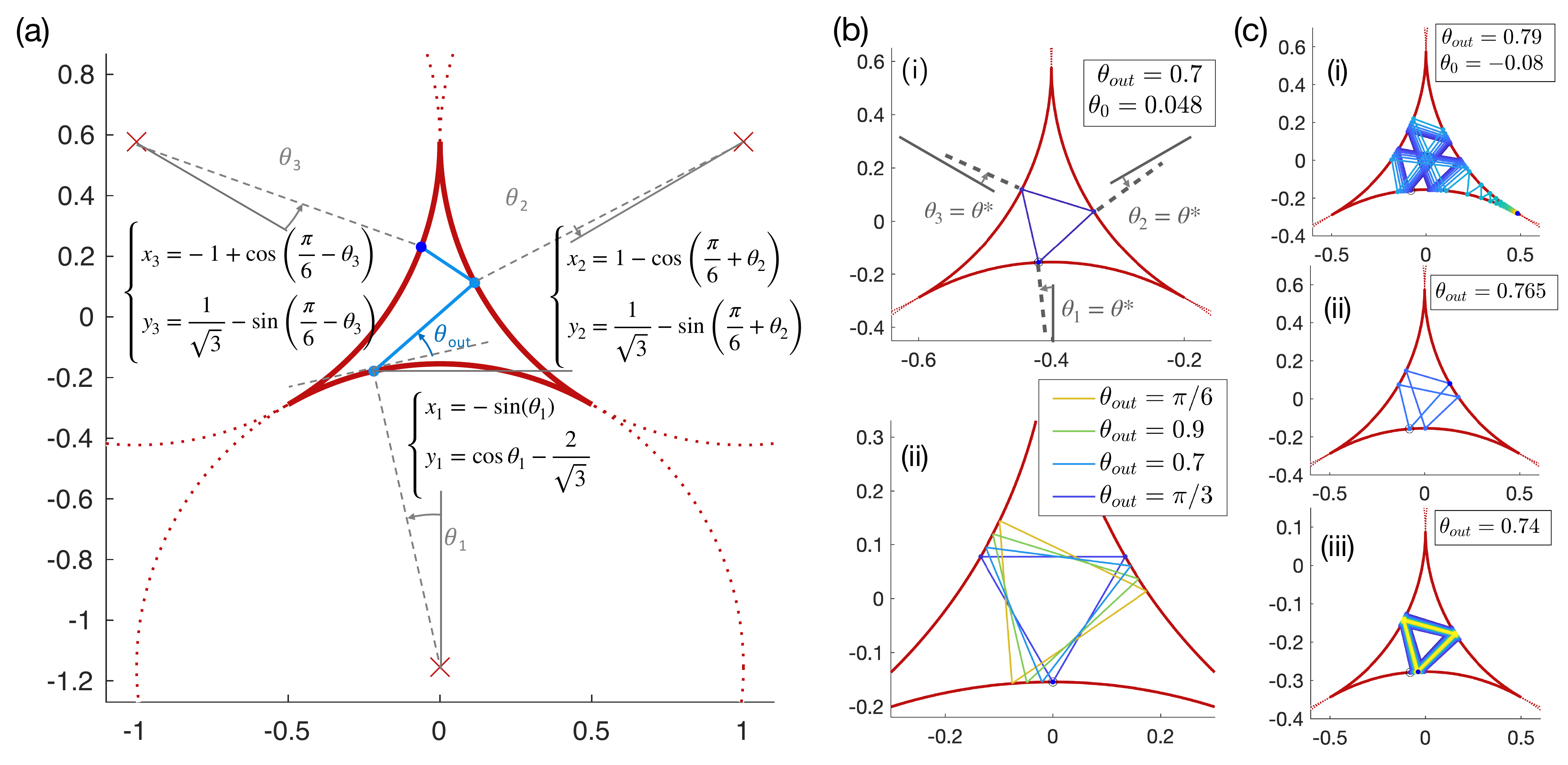}
    \caption{(a) Position of successive contact points on the circles, which we use to derive Eqs.~(\ref{reccontact}) and (\ref{contactleft}). 
    (b) Periodic triangular orbits. 
    (i) Position of the triangle edges as in Eq.~(\ref{eqtriangle}). (ii) Complete range of periodic triangular orbits, with two stable orbits ($\theta_{out} = \pi / {6}, 0.7$) and two unstable orbits ($\theta_{out} = 0.9, \pi / {3}$).
    (c) Trajectories for constant value $\theta_0 = 0.08$ and decreasing values of $\theta_{out}$ showing dynamics ranging from (i) trapping, (ii) six-points periodic orbit, and (iii) asymptotic motion towards the  stable triangular periodic orbit.}
    \label{orbits}
\end{figure}

To do so, we   use the relationship between the Cartesian coordinates of contact points and the orientation of the swimming segment between them. From Fig.~\ref{orbits}a, for a swimmer leaving the bottom circle at $\theta_1$ towards the right one at $\theta_2$, this gives 
\begin{equation}
    \tan ( \theta_1 + \theta_{out}) =  \frac{\sqrt{3} - \cos{\theta_1} - \sin \left(\displaystyle {\frac{\pi}{6} + \theta_2}\right)}{1+ \sin{\theta_1} - \cos\left(\displaystyle\frac{\pi}{6} + \theta_2\right)}.
    \label{reccontact}
\end{equation}

 Similarly, for a trajectory starting on the bottom circle at $\theta_1$ and reaching the left one at $\theta_2$, we get
\begin{equation}
    \tan ( \theta_1 + \theta_{out}) =  \frac{\sqrt{3} - \cos{\theta_1} - \sin \left(\displaystyle{\frac{\pi}{6} - \theta_2}\right)}{1+ \sin{\theta_1} + \cos\left(\displaystyle\frac{\pi}{6} - \theta_2\right)}.
    \label{contactleft}
\end{equation}

There are no two-point periodic orbits for this system, as a swimmer scattering between two circles with constant $\theta_{out}$ goes monotonously towards or away from the corner. The simplest periodic orbits are therefore triangular ones with a swimmer scattering successively on the three circles. 
These triangular periodic trajectories exist when the point reached after one reflection has the same value of $\theta$ (this corresponds to a $2 \pi / {3}$ rotation), and is reached from the same side of the normal. The first condition can be met only when $\theta_{out} <  \pi / {3}$ (see Fig.~\ref{orbits}b).   Besides, the condition for the two sides of the triangle to be on different sides of the normal yields $\theta_{out} > \pi / {6}$ (see Fig.~\ref{mapexit}viii).

We thus find that for $\theta_{out} \in [\pi / {6}; \pi / {3}]$, periodic triangular orbits do exist. We illustrate some of these periodic triangular orbits in Fig.~\ref{orbits}b, including the limit ones. Using $\theta_1 = \theta_2$ in Eq.~(\ref{reccontact}), some trigonometric manipulation leads to the position of the triangle vertices at
\begin{equation}
    \theta_0^* = \arccos{ \left[ \frac{\sqrt{3}}{2} \cos{ \left(\frac{\pi}{6} + \theta_{out}^* \right)} \right] } - \left(\frac{\pi}{6} + \theta_{out}^* \right) 
    \label{eqtriangle}
\end{equation}
The values of the pairs $(\theta_0^*,\theta_{out}^*)$   corresponding to periodic triangular orbits of the system are plotted as a solid white line on Fig.~\ref{mapexit}. 
 Numerically, we observe  that the only stable periodic trajectories are some of these periodic triangular orbits, as illustrated in Fig.~\ref{mapexit}vii. 
 
We now consider their attraction range, and focus on those of the trajectories that tend to periodic triangular orbits. These are included within the central black area in Fig.~\ref{mapexit}, and the smaller black areas for higher values of $\theta_0$. The smaller black zones are trajectories that reach the same value of $\theta$ as the central one after a few reflections (thus leaving the left corner). 

We therefore focus only on the central black area of the phase map.  It is delimited, on the left, for decreasing values of $\theta_0$, by the line for which the swimmer enters the right trapping region  (blue) after one bounce. From the previous section, we know this occurs for $\theta_{0} = \theta_c$ as in Eq.~(\ref{eqtc}). 
The right boundary, for increasing values of $\theta_0$, is the line for which the swimmer enters the top trapping region  (yellow)  after two bounces. This corresponds to taking $\theta_2 = \theta_c$ in Eq.~(\ref{reccontact}) which leads to
\begin{equation}
    \tan(\theta_{out}+\theta_0) = \frac{\sqrt{3} - \sin\left(\displaystyle\frac{\pi}{6} + \theta_c\right) - \cos(\theta_0)}{1 - \cos\left(\displaystyle\frac{\pi}{6} + \theta_c\right) -\sin(\theta_0)}.
\end{equation}
Finally, when increasing the value of $\theta_0$, the limit of the yellow region corresponds to trajectories reaching the left circle first. In the limiting case, the left circle will be reached tangentially, giving rise to $\theta_1 + \theta_{out} = \pi / {3} + \theta_2$. Incorporating this to Eq.~(\ref{contactleft}), we obtain
\begin{equation}
    \theta_1 = \frac{\pi}{6} - \theta_{out} + \arccos \left[\frac{1 + \cos(\theta_{out})}{2} \right].
\end{equation}
 These mathematical boundaries   to the blue and yellow areas of the phase map are plotted as dashed white lines in Fig.~\ref{mapexit}  and match exactly the numerical simulations. 

We obtain that the limit of the attraction region of periodic triangular orbits for increasing values of $\theta_{out}$ is given by six-point periodic orbits. This can be intuitively understood by inspecting  trajectories for a given value of $\theta_{0}$, such as the three trajectories shown in Fig.~\ref{orbits}c. We take a swimmer starting close to the triangular periodic orbit vertex, at $\theta_0 = \theta^* + \delta \theta_0 $ ($ \delta \theta_0 $ is small), and scattering successively on the three circles. We then look for the condition for the swimmer to go towards, or away from, the triangular orbit. Its position after two bounces is given by $\theta_3 = \theta^* + \delta \theta_3$ (see Fig.~\ref{orbits}a for notations). If $\delta \theta_3 < \delta \theta_0 $, then the trajectory is going towards the triangular periodic orbit, and will thus escape the corner; this is  shown in Fig.~\ref{orbits}iii. 
In contrast, if $\delta \theta_3 > \delta \theta_0 $, 
the swimmer goes away from the periodic orbit and ends up reaching the limit for a trapping region, as in Fig.~\ref{orbits}i). The limit case is for the exact equality $\theta_0 = \theta_3$, which is a periodic trajectory with period 6, as in Fig.~\ref{orbits}ii. 
The location of these six-point periodic orbits on the phase map can be found using the $\theta_3 = \theta_1$ relationship, after an intermediate bounce at $\theta_2$. This leads to a system of two equations, namely Eq.~(\ref{reccontact}) together with the same equation inverting the 1 and 2 indices. We can numerically solve for this system and we obtain the dotted line shown on Fig.~\ref{mapexit}. This theoretical prediction is seen to be in excellent agreement with the numerical simulations. 

We finally note the existence of some escaping (black) trajectories on the phase map with $\theta_{out}$ higher than the six-point periodic orbits. This corresponds to trajectories that will end up in a corner at larger times but go slowly away from the six-point periodic orbits. From the previous notations, this means that $\delta \theta_n$ is slowly increasing but the swimmer is not yet trapped after 200 bounces.

\subsubsection{Limits $\theta_{out} \rightarrow {\pi}/{2}$  and $\theta_{out} > {\pi}/{2}$ } 

From the phase map, we see that the largest black region is the one for an increasing value of $\theta_{out}$. Trajectories at high scattering angles appear to be more complex (see illustration in  Fig.~\ref{mapexit}i and v), and for increasing values of $\theta_{out}$, they are less likely to end up in a corner. In contrast with the  central region of trajectories going to periodic orbits, this area does not have a well-defined boundary. In fact, its size can be made to  decrease arbitrarily by increasing the number of bounces in the simulated trajectories. Rather than periodic trajectories, it corresponds therefore to cells spending more time swimming before getting caught in a corner. This is due to the decrease of the trapping region size, which goes to 0 in the limit $\theta_{out} \rightarrow \pi / {2}$ (see Fig.~\ref{critr}b). Swimmers can arrive close to corners and leave them without being trapped, as in Fig.~\ref{mapexit}v,
and in the limit where $\theta_{out} \rightarrow \pi / {2}$, the time before trapping becomes infinite. 

The model could also be extended to consider a swimmer that consistently leaves the wall in the same direction that it arrived from, leading to a constant scattering angle $\theta_{out} > {\pi}/{2}$. This case would of course not be directly relevant  to CR, which is more likely to maintain its swimming direction~\cite{kantsler2013ciliary,ostapenko2018curvature}. 
In that case, the attraction region we studied for $\pi - \theta_{out}$ would now   repel the swimmer away from the corner it entered. 
New orbits of periodicity two, three, four or higher, including stable ones, would occur, and they could be analytically studied using Eqs.~\eqref{reccontact}-\eqref{contactleft}. However, periodic trajectories in this case would not alter the future of the swimmer, which would have a probability  $1$ of escaping regardless of the details of its trajectory.

\subsection{Suppressed escapes: noise and finite algae size}

With the 2D geometrical approach, we have observed the tendency of swimmers to accumulate in corners. However, we have also identified two situations in which   some trajectories do not reach the chamber corners: periodic trajectories for intermediate values of the  scattering angle  and trajectories with a trapping region size going to 0 in the limit $\theta_{out} \rightarrow \pi / {2}$. We now show that incorporating simple additional physical elements, namely noise and the finite  size  of the swimmer,  leads to the suppression  of both these escaping regions.

Firstly, we include noise in the scattering angle. Since we  want to keep a simple geometrical model,   we consider small deviation around a constant $\theta_{out}$ rather than a more realistic distribution of scattering angle as  in the next section. We choose a scattering angle of the form $\theta_{out} = \theta_{out}^{0} + \theta_{noise}$; we take $\theta_{noise}$ to be  normally distributed with a noise strength $\alpha = 0.2$ $\theta_{noise} \sim \alpha \mathcal{N}(0,\sigma_{noise})$ and vary the value of the noise standard deviation $\sigma_{noise}$. For the swimmer to remain in the chamber, we check that it does stay in the interior side of the tangent to the circle after a bounce, and otherwise we redraw $\theta_{noise}$.

 \begin{figure}[t]
    \centering
    \includegraphics[width = 0.85\columnwidth]{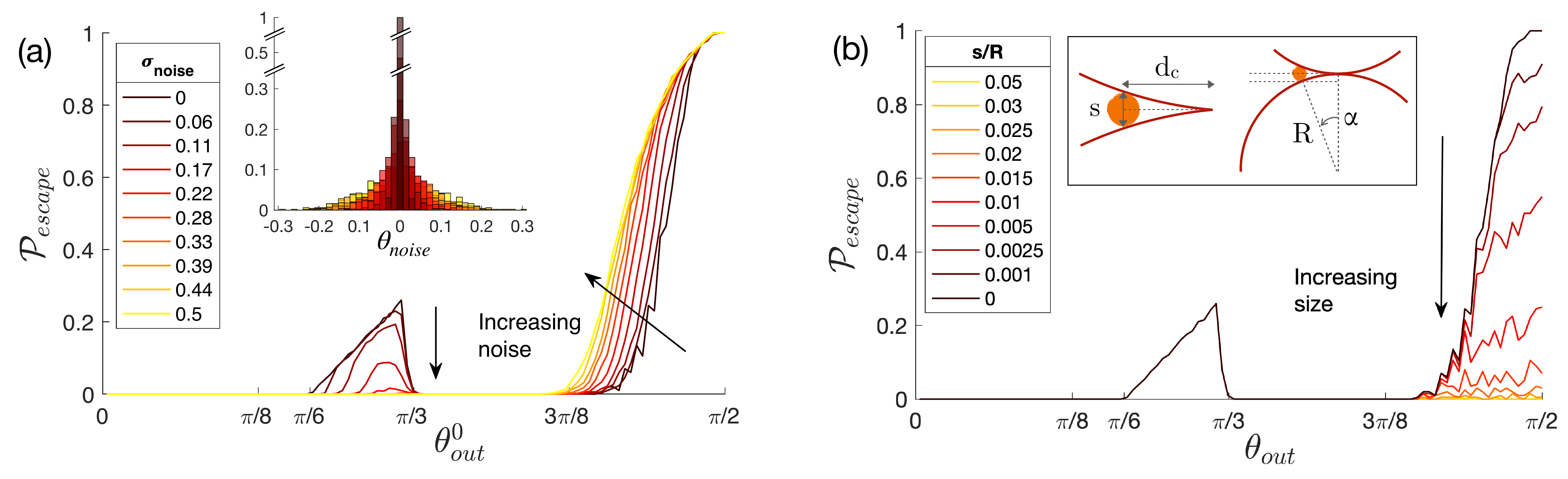}
    \caption{(a) Probability for a model cell to escape (i.e.~to not end up in a corner) after 200 bounces as a function  on the average scattering angle $\theta_{out}^0$ for different values of the noise standard deviation, $\sigma_{noise}$. The inset shows histogram of the corresponding total noise distribution $\theta_{noise}$. 
    (b) Probability for a swimmer to escape after 200 bounces depending on the scattering angle $\theta_{out}$ for different values of the cell size $s$ relative to the circle radius $R$. Inset shows trapping for finite size. Experimental values of the cell size/bubble size ratio in foams are $s/R\approx0.01-0.03$; for these values, escapes for high $\theta_{out}$ is suppressed. }
    \label{pescape}
\end{figure}

We plot the resulting probability for a swimmer to escape as a function of $\theta_{out}$ in Fig.~\ref{pescape}a. This corresponds to integrating the phase map over all values of the initial position $\theta_0$. Without noise, we observe escaping trajectories for both periodic orbits and large scattering angle. Adding even a small amount noise (as shown in the inset histogram of Fig.~\ref{pescape}a) leads to a rapid decrease in the number of trajectories going to periodic orbits. When $\sigma_{noise}$ is  above 0.28 approximately, the periodic trajectories are entirely suppressed. We note that this corresponds to variation in the scattering angle that are much weaker than the  ones measured experimentally~\cite{kantsler2013ciliary}. This suppression of periodic trajectories for noisy values of $\theta_{out}$ is not unexpected, since noise makes the swimmer more likely to go out of the basin of attraction  of a periodic orbit. 
On the contrary, escapes at values of large $\theta_{out}$ appear to become more likely in the case of  noisy scattering angles. 

A closer look at the escaping  trajectories at high values of $\theta_{out}$ reveals that the cells are likely to  approach the corners closely before leaving (see Fig.~\ref{mapexit}v). However, in practice, we expect this to be impossible experimentally, as the swimmer has a finite size.
Taking into account its body size $s$, we may then declare a swimmer to be trapped when it reaches a point where the distance between two walls is smaller than $s$. 
This occurs when the swimmer approaches a corner closer than the critical distance $d_c$. From the inset in Fig.~\ref{pescape}b, 

\begin{equation}
    d_c = R\sin{(\alpha)} \quad \textrm{and} \quad s=2R(1-\cos \alpha)
\end{equation}
and therefore
\begin{equation}
    d_c = R\cos \left[ \arcsin \left( 1 - \frac{s}{2R}\right)\right].
\end{equation}

 With this, we can then plot the escape probability as a function of the value of $\theta_{out}$ for different ratios between the swimmer size $s$ and the surface radius $R$ in Fig.~\ref{pescape}b. Including the  finite  size of the swimmer in our simulations leads clearly to a decrease in the number of escaping trajectories for high values of  $\theta_{out}$.
The relevant experimental parameters for CR in foams and our chambers are $s \approx 10 \si{\mu m}$ for swimmers and $R \approx 1~\si{mm}$ for bubbles. The relevant size ratios $s/R$ are $0.01$ to $0.03$, which correspond to sharp decreases in escaping trajectories ($\mathcal{P}_{escape}<0.2$), though without completely suppressing them. As expected, however, the periodic trajectories are unaffected by finite swimmer size in the absence of noise. 

These simple considerations show that  noise tends to suppress periodic orbits while including the finite size of the swimmers  reduces the number of escaping trajectories at  high values  $\theta_{out}$. By choosing parameters for our model that match experimental conditions (average value of $\theta_{out}$, noise level and algae size), trapping would in fact be predicted to occur for all trajectories. Our purely 2D geometrical model is thus able to explain the main features of the  spatial distribution of CR cells in the experiments, namely their tendency to swim in the corners.

While the geometrical model is able to rationalise corner accumulation, it cannot reproduce quantitatively  the distribution of swimmers in the chamber. There are two main reasons for this:  the trapped swimmers in our model  never escape the corners, and in addition, the geometrical model is very sensitive to the chosen value of scattering angle, $\theta_{out}$, which is nor constant nor normally distributed around a central value in the experiments.  
A full understanding of the probability density functions shown in Fig.~\ref{pdfexp}  requires therefore to incorporate additional elements to our model. In the next section we present a quantitative analysis of the swimming behaviour of the confined algae, and find appropriate laws for our simulations to match the experimental distribution.

\section{Detailed modelling of the  probability density function}
\label{sec:quant}

In this  section, we build on the geometrical model above to reproduce quantitatively the distribution of CR cells in the microfluidic chamber measured experimentally. To do so, we examine  the experimental values of parameters that control the spatial distribution of algae in the chambers. We know from our geometrical model that contact interactions with the walls, including the values of scattering angle, are crucial to understanding the corner accumulation of CR cells. We therefore start by  analysing the experimental distribution of scattering angles and reproduce it in our simulations. We next show that sliding along surfaces also affects the probability of cell presence in the corners. As opposed to the simple approach above, we will now also allow swimmers to escape the corners once they have been trapped. Finally, we go beyond corner trapping and  investigate the accumulation along curved boundaries observed in   our experiments (which was absent from the geometrical model).

\subsection{Scattering angle}

 \begin{figure}[t]
    \centering
    \includegraphics[width = 0.9\columnwidth]{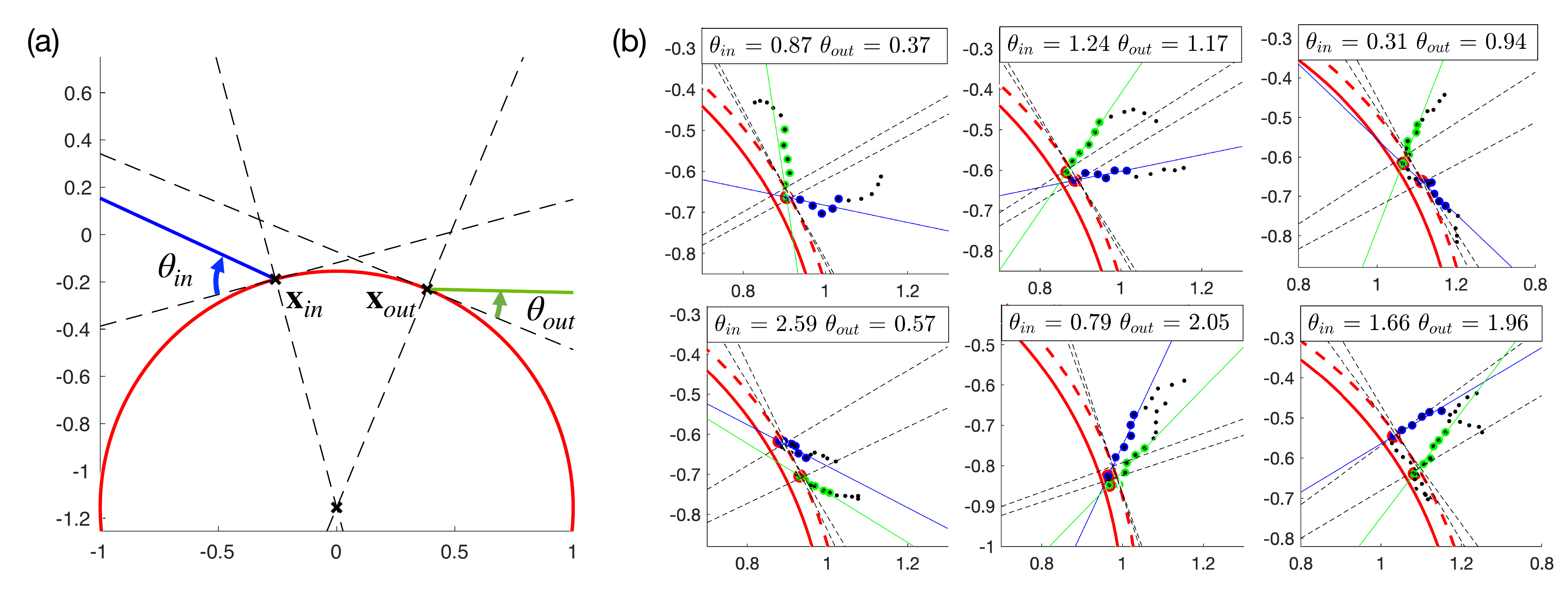}
    \caption{(a) Sketch defining the incident and scattering angles,  $\theta_{in}$  and $\theta_{out}$, respectively.  (b) Illustration of all possible angular configurations of contact events in the $(0,\pi)$ range for both  $\theta_{in}$  and $\theta_{out}$ in the $R = 520~\si{\mu m}$ chamber; the dots are the alga trajectory and the lines show the measured  angles. The dashed line represents the $20 \si{.\mu m} $ surface region with respect to which wall contact is defined. }
    \label{angle520}
\end{figure}

We first focus on the individual contact events with the walls, and in particular on  the orientation of the  CR cells before and after encountering a wall. We measure experimentally the incident and scattering angles,  $\theta_{in}$  and $\theta_{out}$ respectively,  of {\it Chlamydomonas} in our quasi two-dimensional chambers. 
A contact event occurs when the swimmer is on, or very close to,  the surface. We define the surface to be a region of thickness $20~\si{.\mu m}$ near the three theoretical  circular walls.  
This region is shown as a dashed line in Fig.~\ref{angle520} and in green Fig.~\ref{timesurfcor}a. We then measure the distribution of  residence times in the experiments, with results shown in  Fig.~\ref{timesurfcor}b.

 Contact events begin when the cell first enters the wall region of the chamber and end when it leaves it for more than a second. We check that histograms for the  distribution of angles do not change significantly when the minimal take-off time exceeds a threshold of $0.5~\si{s}$. 
This definition enables us to ignore imperfections in the chambers and details of the steric interaction between the cell and the wall. Short consecutive contact events, where the cell bounces on and off the surface while swimming along it, are therefore excluded, and instead considered as continuous sliding along the wall. When comparing our data to past scattering angle measurements~\cite{kantsler2013ciliary,ostapenko2018curvature}, this takes out some small values of $\theta_{out}$ and skews our data towards higher scattering angles.
 Unfortunately, we do not have sufficient precision in our measurement to  describe further the sliding events  in our description. This does not affect our results, which  focus on the probability density function of the algae in the chamber as opposed to details of the contact interactions.

 We also only consider contacts that occur outside the corner region, where steric interactions could instead involve more than one wall at a time so that the measure angles are independent of the local corner geometry and can be readily extended to other microfluidic chambers geometries.
 
 We define the incident and scattering angles, $\theta_{in}$ and $\theta_{out}$, as the angles relative to the tangent at the incidence and departure points, $\bm{x}_{in}$ and $\bm{x}_{out}$, measured using a linear fit to the $0.5~\si{s}$-long trajectories    before and after the contact event (see Fig.~\ref{angle520}a). 
We allow the values of  $\theta_{in}$ and $\theta_{out}$ to be more than $\pi / {2}$ when the cell crosses the normal to the incidence or leaving point, respectively. 
An incidence angle above $\pi/2$ coincides with the algae sliding along the wall in the direction opposite to its arrival at the wall. 
We show in Fig.~\ref{angle520}b examples illustrating all possible contact configurations in the $R=520 ~\si{\mu m}$ chamber, along with the fitted angles. 
 
\begin{figure}[t]
    \centering
    \includegraphics[width =0.85\columnwidth]{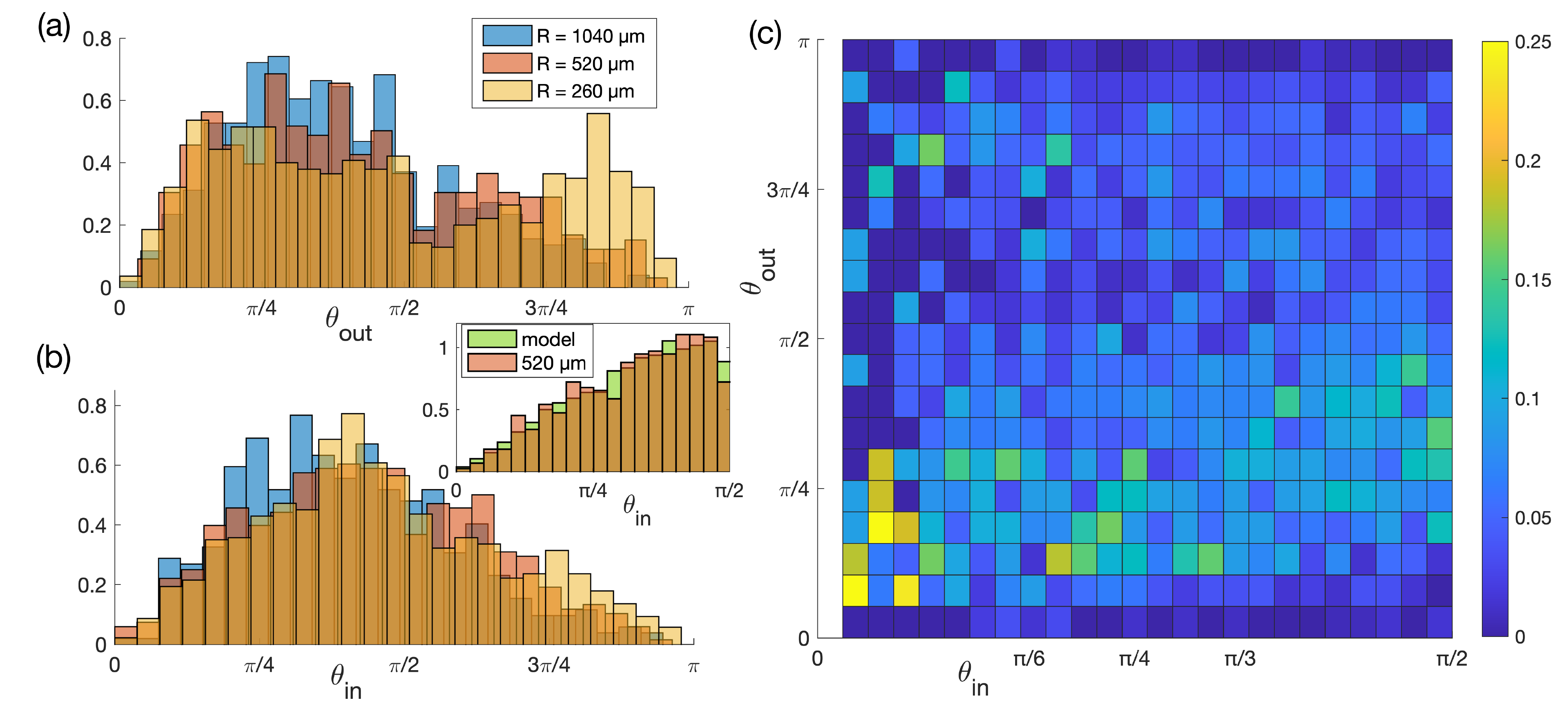}
    \caption{Experimental probability distribution functions for  (a) scattering  and (b) incident    angles for all three chambers (radians) for a total of 2094 contact events. Inset: comparison between experimental and modelled distributions for the incident angle using the experimental  distribution of $\theta_{out}$ from (b). 
    (c) Conditional scattering probability $p(\theta_{out} | \theta_{in})$ measured in our experiments. }
    \label{histoangle}
\end{figure}

We plot the  histogram of $\theta_{in}$  in the three chambers in Figs.~\ref{histoangle}b. We note that there are only small qualitative differences between them. We next  use our results to plot the histogram of  $\theta_{out}$ in  Fig~\ref{histoangle}a.
As mentioned above, our definition of contact events leads to a distribution of scattering angles shifted towards larger values of $\theta_{out}$ (Fig.~\ref{histoangle}a) compared to past work~\cite{kantsler2013ciliary}.

We further note that the   contact angles are not significantly affected by the curvature of the chamber walls.
This can be understood by recalling that the contact interaction dominates the interaction of swimmers  with the walls and thus the scattering angle. Since the radii of the chambers ($R \sim 260 - 1040~\si{ \mu m}$) are large relative to the  size of a CR  (body radius, $ s \sim 5~\si{ \mu m}$ with flagella $ \sim 10~\si{ \mu m}$), there is a clear separation of length scales.

The distribution of incidence angles is maximal around $\pi / {2}$ and decreases continuously towards $0$ and $\pi$ (Fig.~\ref{histoangle}b). Is it predominantly set by local cell-wall interactions or by the global chamber geometry?  To understand the origin of the distribution triangular shape, we compare it with the same histogram obtained from the  geometrical model from Section 3. We note that in that case, we always had that $\theta_{in} < \pi / {2}$ since we did not include the displacement of the cell along the surface. 
We find that the distribution of $\theta_{in}$ obtained using a constant value of $\theta_{out}$ does depend on the value of $\theta_{out}$. However, if we draw the value of the scattering angle from the experimental distribution shown in 
 Fig.~\ref{histoangle}a, we obtain a histogram qualitatively identical to experiments, as shown in the inset of (b). Using other physical distribution of scattering angle~\cite{kantsler2013ciliary} produces the same result. We have thus shown that the distribution of incident angles is  dominated by the geometry of the system described in the previous section rather than by hydrodynamic interactions of the swimmers with the wall. 
 
As a difference with the incident angle, the distribution of $\theta_{out}$ in Fig.~\ref{histoangle}b shows  two distinct regions, $p_{\theta_{out}<\pi / {2}}$ and $p_{\theta_{out}>\pi / {2}}$, with a peak  for $\theta_{out}< \pi / {2}$. Overall, the  probabilities for $\theta_{out}$ to be below or above $\pi / {2}$ are approximately $66\%$ and $34\%$. 
We then consider the conditional scattering probability distribution for  $\theta_{out}$ given the incident angle, i.e.~$p(\theta_{out} | \theta_{in})$, which has been investigated close to straight walls~ \cite{kantsler2013ciliary} and to pillars~\cite{contino2015microalgae}. In order to use our results for simulations and for comparison with existing literature, we assume $\theta_{in} \in [0, \pi / {2}]$  and if $\theta_{in} > \pi / {2}$, we redefine $\theta_{in}$ to be its complement to $\pi$.  The resulting plot for the 2094 contact events in all chambers is shown on Fig.~\ref{histoangle}c. It appears that the values of $\theta_{in}$ and $\theta_{out}$ are not independent;  for a given value of $\theta_{in}$ we mostly observe two peaks for $\theta_{out}$, the highest below $\pi / {2}$ and the second one above it.  From this point on, we will use this measured angle distribution in the chambers in order to obtain an accurate spatial distribution of swimmers in our simulations. 

\subsection{Time at the wall}

The second important feature of the experimental distribution of CR in the microfluidic chambers is the increased concentration of swimmers close to the boundaries. 
This concentration has been described in past work in the case of a curved surface based on purely steric interactions and 
noise~\cite{ostapenko2018curvature} and it has been experimentally measured~\cite{kantsler2013ciliary}, but both studies  considered only steric (not  hydrodynamic) interactions. We will first consider algae in contact with the wall and measure the time they spend swimming along it. 

\begin{figure}[h!]
    \centering
    \includegraphics[width = 0.9 \columnwidth]{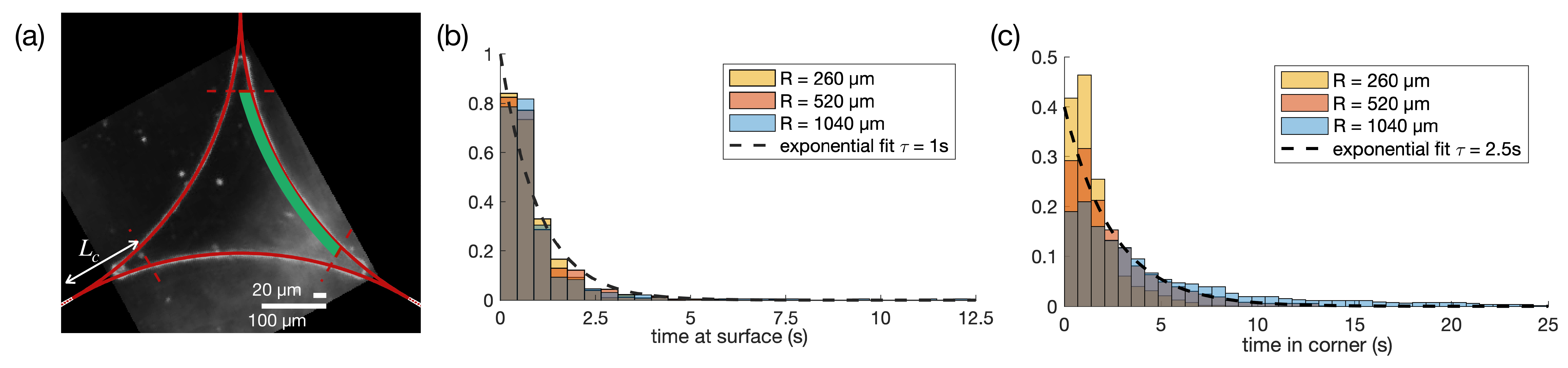}
    \caption{(a) Definition of the `surface' and `corners' to quantity residence times in the experimental  chamber of radius $R=520~\si{\mu m}$. The right surface region is depicted in green and the small scale bar of $20~\mu$m shows its size while $L_c$ is the corner extension. (b)  Probability density function for the duration of surface contact events in the three chambers; the dashed line shows an exponential fit with characteristic time $1~\si{s}$. (c) Probability density function of the time spent by the cells in a corner, together with an exponential fit with a characteristic time scale of $2.5~\si{s}$ for the middle $R = 520~\si{\mu m}$ chamber.}
    \label{timesurfcor}
\end{figure}

A contact event occurs when the swimmer is on, or very close to, the surface. As when measuring scattering angles, we define the surface to be a region of thickness $20~\si{.\mu m}$ near the three theoretical  circular walls (Fig.~\ref{timesurfcor}a) and measure the distribution of  residence times in the experiments, with results shown in  Fig.~\ref{timesurfcor}b. The
 average duration of residence near the wall  is $1~\si{s}$. We observe no significant change for the duration of contact between chambers, except some longer contacts for the $1040 ~\si{\mu m}$ one (up to $35~\si{s}$) due to  dust near the boundary that locally trapped cells for long periods. During this contact, the cells  move along the surface for an average of $31~\si{\mu m}$ ($21~\si{\mu m}$ in the $260~\si{\mu m}$ chamber, $38~\si{\mu m}$ in the $520~\si{\mu m}$ chamber and $45~\si{\mu m}$ in the $1040~\si{\mu m}$ one). Interestingly, the contact with a wall is not a memory-less event, as steric interactions do take time to cause reorientation of the algae~\cite{kantsler2013ciliary}. Nevertheless,  the exponential fit shown in Fig.~\ref{timesurfcor}b is able to capture the experimental distribution. Any further study  of  contact interactions with  surfaces would require including the detailed shape of the swimmer~\cite{chen2020shape} and  the interactions between the walls and  the flagella~\cite{kantsler2013ciliary}, both of which are beyond the precision of our experiments. 

\subsection{Slowing down when approaching the wall}

\begin{figure}[t]
    \centering
    \includegraphics[width = 0.55\columnwidth]{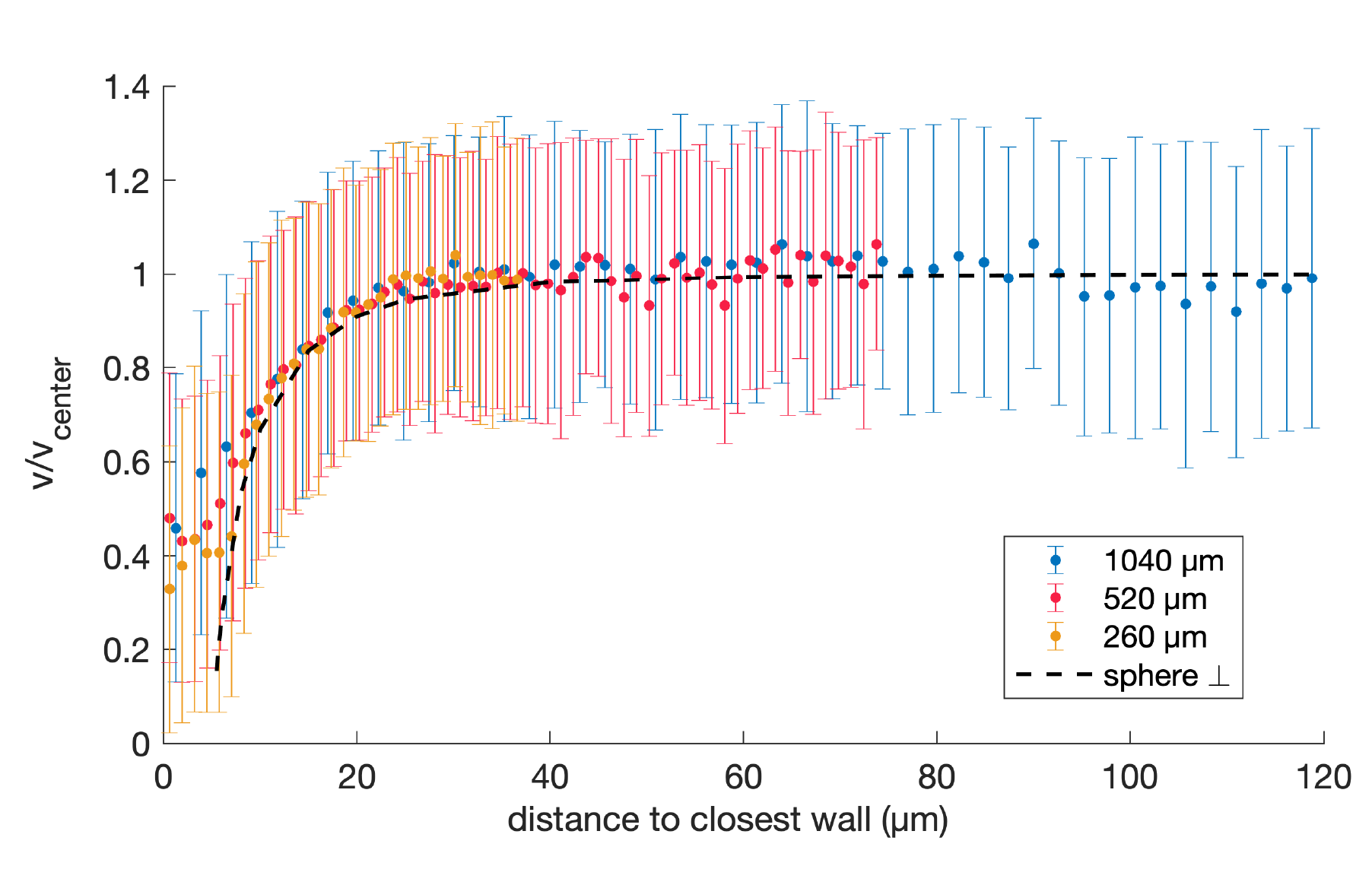}
    \caption{Average swimming speed of   {\it Chlamydomonas} cells as a function of their distance to the nearest wall (in microns), rescaled by the speed in the  centre of the  chamber, for the three chambers. The error bars show the standard deviation, which is of similar size for individual trajectories (individual cells experience significant speed variations, sometimes for several seconds). The dashed line shows the speed  of a forced sphere of diameter $10~\si{\mu m}$ confined between to walls separated by a distance $20~\si{\mu m}$ when approaching the third wall perpendicularly, as obtained from numerical simulations.}
    \label{speedwall}
\end{figure}

 We next address  long-range interactions between the microorganisms and the walls (i.e.~that don't have a steric origin). 
 We measure the velocity of the CR cells as  a function of  their distance to the nearest wall. To do so, we approximate the channel surface by three mathematical circles of radius $R$, and measure the distance of the swimmer to the closest circle centre (Fig.~\ref{timesurfcor}a).  Note that this method introduces some uncertainty in the distance measured  since the channel walls in the experiments are not perfectly regular, leading to an error of order $\pm 5~\si{\mu m}$; depending on the local shape of the surface, we can thus occasionally find some cells swimming as close as $1~\si{\mu m}$ from the theoretical boundary. We include only algae that interact strongly with a single  boundary at a time, outside the corner regions. Furthermore, we only take into account trajectories for which the speed in the middle of the chamber, denoted by $v_{\rm center}$, is at least $2/3$ of the average speed in the middle for all trajectories in a given chamber (keeping 25 out of 30 trajectories). This method eliminates CR cells that could have deficient swimming. The average swimming speed close to the centre of our chambers is found to be $111 \pm 33~\si{\mu m}.\si{s^{-1}}$ for the $260~\si{\mu m}$ chamber, $114 \pm 31~\si{\mu m}.\si{s^{-1}}$ for the $520~\si{\mu m}$ chamber and  $123 \pm 27~\si{\mu m}.\si{s^{-1}}$ for the $1040~\si{\mu m}$ one.

With this methodology, we plot in Fig.~\ref{speedwall} the swimming speed of the cells normalised by their speed far from the walls, $v_{\rm center}$. We observe an almost perfect collapse of the plots not only for individual experiments  but also for all three chambers. 
It is notable that the swimming speed undergoes a sharp decrease when the cells approach walls. The decrease starts approximately $40~\si{\mu m}$ away from the closest surface and is therefore not caused by steric interaction.

\begin{figure}[h!]
    \centering
    \includegraphics[width = 0.8\columnwidth]{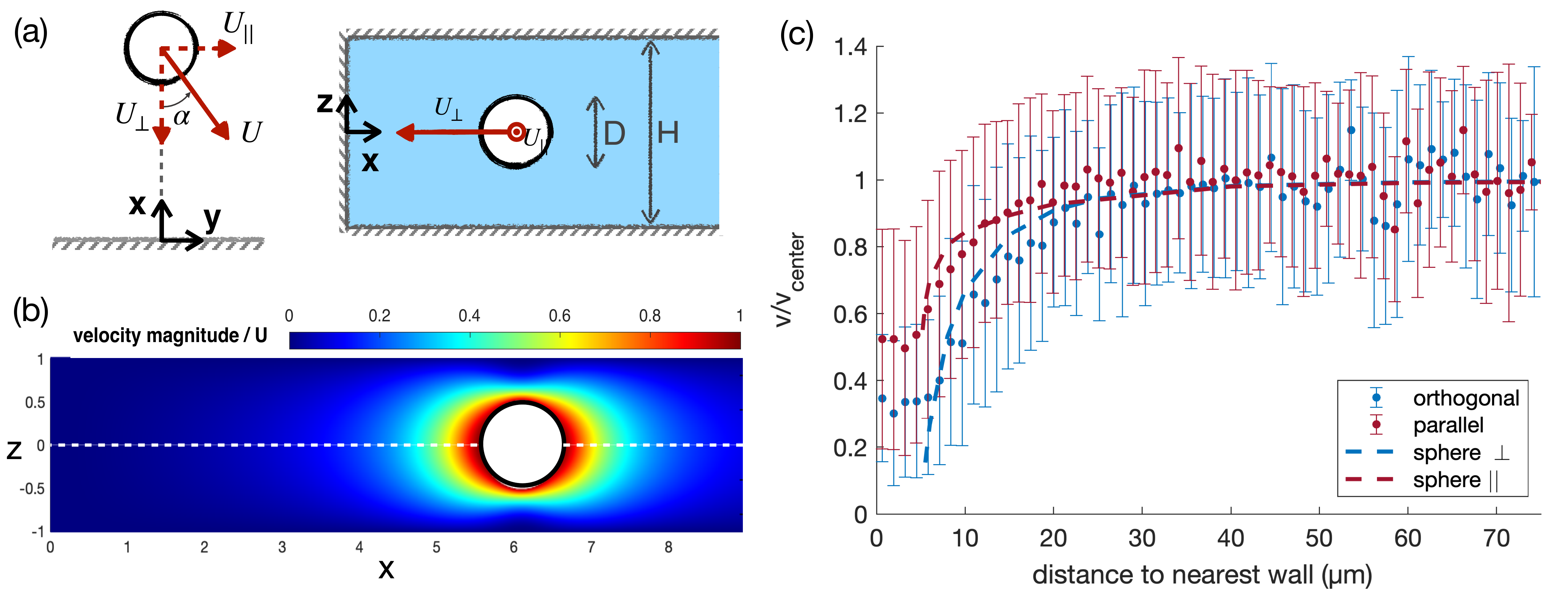}
    \caption{Computational model for the drop in swimming speed for confined algae  swimming close to a wall at $x=0$. 
    (a): Sketches defining the angle relative to the wall, $\alpha \in [0; \pi ]$, the parallel $U_{\parallel}$ and perpendicular  velocities $U_\perp$ and the parameters in the numerical simulations. 
    (b) Example of computed velocity magnitude in the chamber in the orthogonal case (i.e.~motion toward the wall at $x=0$), with $D=H/{2}$ and  a distance $x = 60~\si{\mu m}$ from the wall. 
     (c) Comparison between the experimental swimming velocities (in the $R=520~\si{\mu m}$ chamber) and those obtained numerically (dashed lines)  as a function of the distance to the wall $x$ for  orthogonal motion (orientation angles in the range $\alpha < \pi / {8}$ or $\alpha > \pi - \pi / {8}$) and parallel motion (angles in the range $ \pi / {2} - \pi / {8} < \alpha < \pi / {2} + \pi / {8}$). In the numerical simulations, we take $D=10~\si{\mu m}$ and $H=20~\si{\mu m}$. }
    \label{spherespeed}
\end{figure}

In order to understand the decrease in speed of the swimmer when approaching the wall, we perform numerical simulations using COMSOL Multiphysics$^{\mbox{\tiny \textregistered}}$, with notation shown in Fig.~\ref{spherespeed}a. We consider a sphere of diameter $D$ in a confined domain  of thickness $H$ (so no-slip surfaces are  located at $z=\pm H/2$) in the presence of a third no-slip wall at $x=0$. The centre of the sphere is located at a distance $x=d$ from the left wall. The computational domain is closed with  three rigid surfaces located far from the sphere at $x=25D$ and $y=\pm 25D$ that  do not affect the flow. We non-dimensionalise the geometry with the radius of the sphere, $D/2$.

 We consider two setups; in the first one the sphere is moving towards the wall at $x=0$ and in the second one parallel to it. In both cases the sphere has a unit velocity. The problem is solved using the Fluid Module of COMSOL, with  tetrahedral mesh of about 966k elements. An example of a velocity magnitude profile for a sphere swimming towards the wall is illustrated in Fig. \ref{spherespeed}b. We use the simulations to compute the drag force on a sphere,  denoted by  $F_i$, with the notation $i=\perp,\parallel$ used to denote  motion orthogonal and parallel to the surface at $x=0$, respectively. To simulate the effect of an organism pushing with constant force, rather than having constant velocity, we may invoke the linearity of  Stokes flows and rescale
 at each point    the force on the sphere to be that in the centre of the domain, $F_{\rm center}$, so that the speed of the sphere is given by $U_i=F_{\rm center}/F_i$ with  $i=\perp,\parallel$ (see Fig.~\ref{spherespeed}).
 
Modelling a CR as a sphere dragged by a constant force turns out to lead to a good agreement with our measurements of swimming speed vs distance to wall, as demonstrated in Fig.~\ref{speedwall} showing both the correct range  of hydrodynamic interactions between the surface and the swimmer and the magnitude of the decrease.  
We note that the velocity drop in the simulations is smaller when the sphere is moving parallel to the walls; this is also in good agreement with our experiments with the confined algae, as we show in Fig.~\ref{spherespeed}c by plotting the swimming speed  only for a range of orientations $\alpha$ relative to the nearest surface (see details in figure caption). 

A  real swimming cell is, of course, not subject to a constant force, but is instead free-swimming. Our  modelling approach focuses therefore only on the hydrodynamic interaction between the confined cell body and the walls, ignoring any effects arising from the smaller flagella. The good agreement between numerics and experiments allows us to confirm, as posterori, that these wall-body interactions govern the long-range interactions in the experiments.

Experimentally, we note an asymmetry between the speed of algae swimming away from the wall and towards the wall, the latter being smaller. This feature cannot be reproduced from the sphere model, in which the two cases are identical (a consequence of reversibility of Stokes flows). Recent experiments \cite{jeanneret2019confinement,mondal2020inversion} have shown that the flow field around a confined {\it Chlamydomonas} is dipolar (two-dimensional source dipole).  Furthermore, the impact of the dipolar flow field created by the algae body  increases with confinement relative respect to the propulsive forces from the flagella~\cite{jeanneret2019confinement}.  
 It was noted in past work  that when fitting the experimental flow of a confined {\it Chlamydomonas} to a dipolar model, the source dipole should be positioned in front of the algae rather than in the middle of the cell body~\cite{jeanneret2019confinement} . This additional parameter leads to a shift in the distance from the wall at which the velocity decay occurs for forward and leaving algae. We measure this shift to be $\sim 5\si{.\mu m}$ in our experiments, consistently with a characteristic CR size.

\subsection{Trapping in a corner}

When  a {\it Chlamydomonas} cell reaches a corner of the chamber, we observe experimentally that it stays trapped there longer than at a single wall. This is expected due to the geometry of the chamber and the results from our geometrical model.  
We plot the distribution of trapping times for the  cells in Fig.~\ref{timesurfcor}c. Since the corners of the PDMS chamber are truncated rather than acute, they are not really  biologically relevant to the case of foam structures and there is  need to focus on the details of the cell behaviour in a corner.

Assuming that escaping the corner are events that occur continuously and independently from one another at a fixed rate, we expect the probability density function of the residency time in a corner to be an exponential law. We find a good fit, shown in Fig.~\ref{timesurfcor}c, for this exponential distribution with  characteristic times in the three chambers given  by  $\tau_{260} = 1.5~\si{s}$,  $\tau_{520} = 2.5~\si{s}$, and 
$\tau_{1040} = 3.5~\si{s}$. The longer residency times in larger chambers is due to the local corner geometry and in particular the increasing sharpness of the corner. The corner size $L_c$ is approximately $22\si{.\mu m}$ for  the chamber with $R=260\si{.\mu m}$, $17\si{.\mu m}$ for $R=520\si{.\mu m}$ and $16\si{.\mu m}$ for $R=1040\si{.\mu m}$. 

\subsection{Final model for the cell probability density function}

Using our analysis of the different aspects of the swimming behaviour of CR in a  microfluidic chamber, we can modify the model to  reproduce the probability density function for the position of swimming algae in a microfluidic chamber. We use methods similar to the ones described in the geometrical model of Section~\ref{geomodel} but  we fit the model parameters with the measured quantities from the experiments, thereby incorporating the parameter distributions directly from the   data.  Instead of a binary trapped vs escaped fate for the swimming trajectories, our modified simulations   lead now to a spatial probability density function for the swimmers in the chamber.

\subsubsection{Scattering in model}
We first  use a  distribution of scattering angles $\theta_{out}$ drawn from experiments. 
We  use Fig.~\ref{histoangle}c to draw the scattering angle after each contact event. Using the knowledge of $\theta_{in}$ in a $0.1~\si{\radian}$ interval, we draw $\theta_{out}$ using the corresponding experimental distribution. Neglecting the correlation between incidence and scattering angle does not affect the final value of the  probability density function. 

\subsubsection{Sliding in model}

We next incorporate  sliding during contact events  by moving the point at which the swimmers leaves the wall from  $\theta $ to a new value of $ \theta + \delta \theta$. For each bouncing event, we first draw the scattering angle $\theta_{out}$ when the swimmer reaches the surface at $\theta$. A value $\delta \theta$ is then drawn randomly from an exponential distribution with mean $0.35~\si{\radian}$ to match the sliding measured from experiments in the $R=260~\si{\mu m}$ chamber. The swimmer then leaves the wall at $ \theta + \delta \theta$ with angle $\theta_{out}$ measured relative to the tangent to the circle. The speed of the swimmer during this displacement is fixed to $45~\si{\mu m.s^{-1}}$, and its distance to the wall is drawn uniformly in the interval $[0,10~\si{\mu m}]$.  
Taking into account the sliding along the wall increases naturally the trapping probability, with a simple  geometrical explanation: the displacement along the wall is equivalent to taking a smaller scattering angle $\theta_{out}$ upon leaving from the impact point at $\theta$. Geometrically, it  corresponds to a shift of the distribution of scattering angles towards lower values $\theta_{out}$. Ignoring sliding but taking instead the angular distribution from previous work~\cite{kantsler2013ciliary}, which has lower scattering angles,  would thus result in a similar probability density function. 

\subsubsection{Slowing down near walls in model}

We next include the  change in the swimming velocity of the cells as they approach a wall,   using the fit of Fig.~\ref{speedwall} with a minimum velocity taken to be $45~\si{\mu m.s^{-1}}$. While this effect does change slightly the presence probability for the swimmers, it turns out to be negligible compared to the corner accumulation in our simulations. This demonstrates that in the experiments the mechanism for surface accumulation is primarily the combination of  steric interactions with swimming along the surface during contact events, as opposed to long-range hydrodynamic effects, in agreement with previous studies~\cite{kantsler2013ciliary,ostapenko2018curvature}. 

\subsubsection{Leaving corners in model}

Finally, and  as opposed to what was done in Section~\ref{geomodel}, the simulations now allow cells to  leave a corner after being trapped in it for a given time. This time is exponentially drawn from experimental characteristic times that vary  with the size of the chamber and the sharpness of the corners (see Fig~\ref{timesurfcor}c). To match the corner size in the experiments, we take the  corners in our model to be $0.3R$ long (see Fig.~\ref{pdfexp}). 

\subsubsection{Predictions of new model}

 \begin{figure}[t]
    \centering
    \includegraphics[width = 0.85\columnwidth]{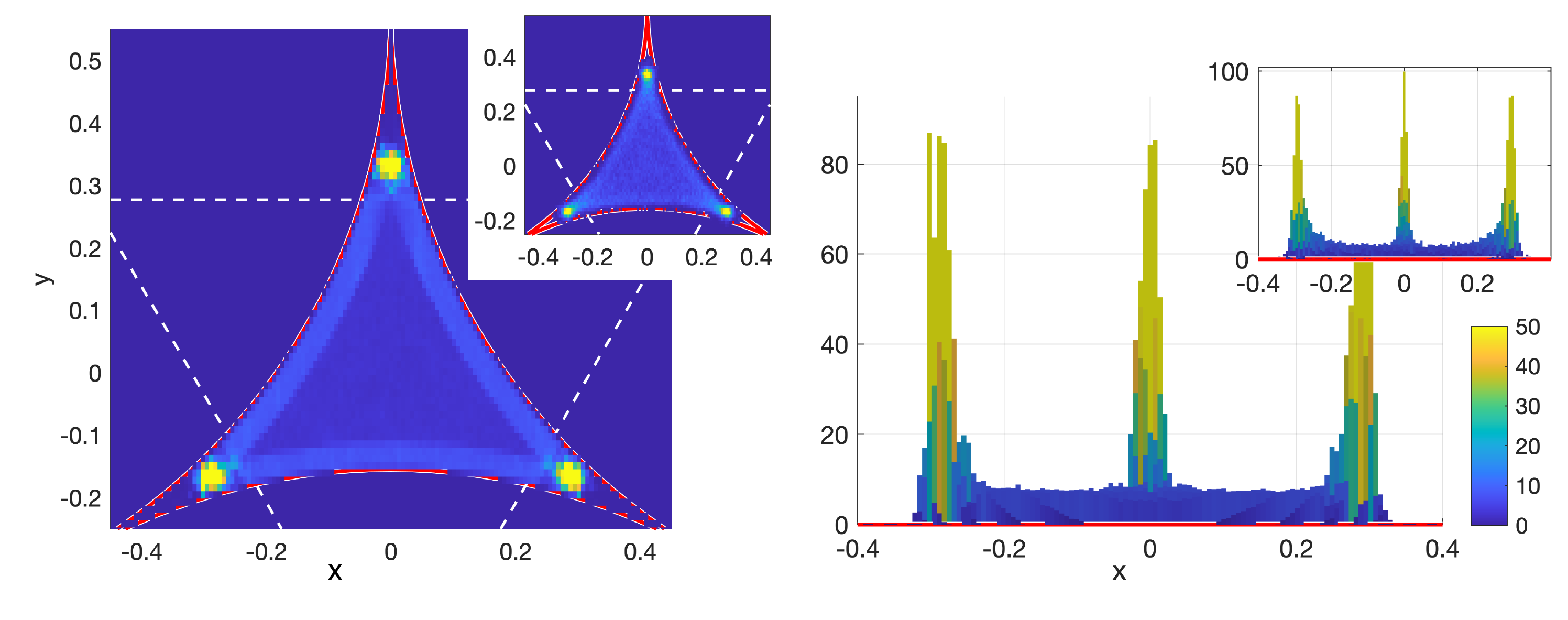}
    \caption{Probability density function of a swimmer from the final model, with parameters taken from experiments in the $R = 260~\si{\mu m}$ chamber. Top: top view; Bottom: side view.  Insets: experimental probability density function from the $R=260~\si{\mu m}$ chamber, for comparison.}
    \label{theopdf}
\end{figure}

With all these modifications included, we can now compare directly the experimental pdf with that resulting from our model. We illustrate in Fig.~\ref{theopdf} the prediction for the distribution from our simulations based on the parameters measured for the intermediate chamber, i.e.~with $R = 260~\si{\mu m}$. The agreement between the results of the model and the experiments is excellent. 

In the simulations, the swimmers end up spending 29\% of their time trapped, which matches well with the 32\% measured experimentally (see Fig.~\ref{pdfexp}a). 
Carrying  out the same simulations with parameters from the other two chambers, we obtain trapping 22\% and 14\% of the time in the $520~\si{\mu m}$ and $1040~\si{\mu m}$ chambers, respectively, both of which compare well with the figures of 29\% and 14\% from our  experiments. 
Remarkably, although the model does not of course reproduce exactly the trajectories for cells swimming in the microfluidic chamber, which would be noisy~\cite{ostapenko2018curvature,wan2014lag}, the modelling illustrated in Fig.~\ref{theopdf} successfully identifies the main ingredients governing the spatial distribution of the microswimmers in the chamber. 

In our model, the interactions between a single wall and a swimmer are all encoded in the distribution of scattering angles, sliding along the wall, and long-range hydrodynamics. These are local properties of a confined system and are thus independent of the chamber geometry and in particular of chamber size. Their distribution could therefore be used directly in different chamber geometries with the same swimming cells. 
In contrast,   trapping  in corners is exponentially distributed with a characteristic time  that depends on the cusp steepness. Extending our model quantitatively to different geometries would therefore require a new calibration of the corner trapping time, or at least an extrapolation of our   measurements. 
Other microswimmers, such as different strains of {\it Chlamydomonas} or swimming bacteria, are expected to have different steric interactions with boundaries~\cite{kantsler2013ciliary}. This would result in different behaviours when bouncing and swimming along walls. Adapting our model to those would then require measuring the distributions of the elements that we have shown   dominate the  spatial distribution of swimmers and incorporating them in our  analysis.

\section{Conclusions}
 
In this paper, motivated by the fate of algal cells in foams, we investigate the spatial distribution of motile CR cells in microfluidic chambers shaped as cross-sections of Plateau borders. We observe that the cells are more likely to be found in the corners of the channel, where they spend about a third of their time. To explain this tendency to swim in the chamber corners, we first develop a  geometrical (billiard) model with reflection laws adapted to the case of swimming microorganisms. Namely, we consider trajectories of model cells bouncing on walls with constant scattering angles to mimic steric interactions with a boundary. The cells swim otherwise  in straight lines inside a domain bounded by three disks, to represent the cross-section of foam Plateau borders. We find that most trajectories end up converging towards a cusp of the domain. We quantify corner accumulation by analysing the phase space diagram of all possible trajectories, which are fully characterised by their initial position and constant scattering angle. In particular, we show that small scattering angles lead to corner accumulation, while the trapping time increases at large scattering angles. We also discover some periodic trajectories, which are suppressed when including noise in the choice of the scattering angle. Our model shows therefore that corner accumulation has a geometrical origin. In particular, cusps create attracting trapping regions for swimmers bouncing on the walls of closed system with acute angles. This is a generic result relevant to many different geometries, which is therefore likely to be significant in a wide variety of complex environments beyond the microfluidic chambers presented here. Corner angles  and  cusps are thus likely to cause an accumulation of microswimmers regardless of the details of hydrodynamic or steric interactions with their environment. 

We next develop a more detailed model based on data from experiments to explain and reproduce quantitatively the location of the CR cells in the chambers. From our geometrical model, we know that the value of the scattering angle, and more generally  wall interactions, control the location of the swimmers. We define and measure experimentally three main elements in a contact event: the incident angle $\theta_{in}$, the scattering angle $\theta_{out}$, and the distance during which CR swims along the wall without leaving it for more than $1~\si{s}$. While the value of $\theta_{in}$ stems from the geometry of the system, $\theta_{out}$ is found to have a complex distribution, which we use as an empirical law for our simulations. We observe that $\theta_{in}$ and $\theta_{out}$ are slightly correlated but the scattering distribution is very noisy and this correlation does not affect the result of our simulations. The time spent by the cell swimming along the wall does explain the boundary accumulation that we observe experimentally, and enhances corner accumulation. Geometrically,  sliding along the walls is equivalent to a shift of the scattering angle distribution towards lower $\theta_{out}$. We also observe that   walls exert a long-range hydrodynamic influence on the CR cells, which slow down when swimming close to a wall. This  can be reproduced accurately by taking into account the hydrodynamic drag acting on the body of the confined algal cells in the 2D chambers. 
The swimming elements we incorporate in our model, notably the scattering angles and sliding distribution that describe cell-wall interactions, are local and robust to change in the geometry of the confined chamber.
Including finally the distribution of time the swimmers spend trapped in the corners allows us to obtain a  complete model that quantitatively reproduces the probability density function of CR cells in the microfluidic chambers. 
This is the only quantitative element of our model that is geometry-dependent: steeper corners lead to longer trapping times, and a new calibration or fitting from our own measurements is needed to extend the model quantitative predictions.  

We describe the swimming behaviour of {\it Chlamydomonas} algae through local swimming properties, and thus expect our full quantitative model to be valid for other confined geometries. However, the interaction between a microorganism and a boundary is organism-dependent~\cite{kantsler2013ciliary}, and when using different strains of {\it Chlamydomonas} or other swimmers, scattering angle and sliding distributions should be experimentally determined.

This work was initially motivated by an experimental study of motile {\it Chlamydomonas} algae remaining trapped in a foam  draining under gravity~\cite{roveillo2020}. We have shown here that the local geometry of the  channels in the foam is likely to play a decisive role in the spatial distribution of motile CR and therefore in their trapping. 
Indeed, if the motile cells spend a large fraction of their time stuck in the  cusps of the Plateau Borders, this will slow down their escape and  effectively retain them in the foam. 

To  address fully the problem of trapping of planktonic microorganisms in foams, the third dimension of the Plateau borders should of course  be taken into account. The first natural extension of our geometric approach to an elongated channel would be to incorporate a new swimming component in the third direction. The 2D model would then be a projection of the 3D one. Scattering angle distribution and wall sliding properties might be affected, and would need to be experimentally measured close to an unconfined wall. However, the results would be qualitatively identical and quantitatively similar to those presented above. Crucially, corner accumulation is a robust feature of the three-circle geometry regardless of the local details of swimmer-wall interactions. Additional 3D kinematics, including rebounds on the top and bottom walls, would need to be included  to describe algae swimming in thicker chambers. 

However, we expect new features that are absent in 2D to play a substantial role in a 3D Plateau borders, predominantly flow and gravity.   
Gravity is known to bias the movement of {\it Chlamydomonas} algae through gyrotaxis, encouraging cells to swim upwards. Gravity also leads to a slow sedimentation of the algae,  which are slightly heavier than water. In this case, the orientation of the channel relative to the vertical direction would have an additional influence on the accumulation of swimmers, and would introduce differences in accumulation between horizontal and vertical borders, as well as an asymmetry between the corners. A similar effect could be induced by the responses of the cells to other cues, including phototaxis (light).

Furthermore, for a foam draining under gravity, motile swimmers will interact with the bulk flow through both advection and   reorientation due to shear. 
The effect of the flow will vary   in the different regions of the foam channels. In the central area, shear coupled to gyrotaxis could concentrate swimmers in the centre. However, the liquid speeds and hence shear rates in the Plateau borders will be small in the corners compared to the central region. This could then contribute to corner accumulation: advection is slower there, and algae already retained in the corners have less chances to escape by shearing. 
We expect a competition between the geometric  effects of swimming highlighted in our paper and the interactions with the flow.  
In contrast, for non-motile cells or non-Brownian tracers, the dominant effect will be passive advection with the draining flow, as seen experimentally~\cite{roveillo2020}. 

Our present study could be extended to incorporate these elements and investigate which increase  the retention of algae in the draining foam.

The method we develop in our paper, combining a purely geometrical billiard model with non-elastic bouncing with experimental data to reproduce accurately scattering and wall motion, could be applicable to other geometries and organisms. Since microswimmers often inhabit complex environments, e.g.~soil or porous media and biological tissues, the local features of their habitats are expected to influence their ability to move, escape local asperities and access resources. Singular features such as corners are expected to be important for cell motility.
In particular, cusps create local traps for swimming cells and are likely to retain them. We hope that our experimentally-driven approach to microorganism billiards will be used to complement existing models of swimmers in complex media, such as active Brownian particles and models with full  hydrodynamic interactions, helping uncover the impact of  local geometry on the dynamics of swimming cells.

\section*{Conflicts of interest}
There are no conflicts to declare.

\section*{Acknowledgements}
This project has received funding from the European Research Council (ERC) under the European Union's Horizon 2020 research and innovation program (grant agreement 682754 to EL). F.E. thanks the CNRS for support from MITI 2019 and 2020.

\nocite{*}
\bibliographystyle{abbrvnat}
\bibliography{main}

\end{document}